\documentclass[a4paper,12pt]{article}

\usepackage[centertags]{amsmath}
\usepackage{amsmath}
\usepackage{braket}
\usepackage{amsfonts}
\usepackage{amssymb}
\usepackage{amsthm}
\usepackage{newlfont}
\usepackage{bbm}

\theoremstyle{plain}
  \newtheorem{theorem}{Theorem}[section]
  
  \newtheorem{proposition}[theorem]{Proposition}
  \newtheorem{lemma}[theorem]{Lemma}
\theoremstyle{definition}

\theoremstyle{remark}

\numberwithin{equation}{section}

\DeclareMathOperator{\Tr}{Tr}
 
\DeclareMathOperator{\Prob}{\boldsymbol{Prob} }

 \let\be=\beta \let\de=\delta \let\ep=\epsilon
   
\let\ka=\kappa \let\la=\lambda \let\om=\omega 
\let\si=\sigma

  \let\La=\Lambda \let\Om=\Omega

\newcommand{\caA}{{\mathcal A}}
\newcommand{\caB}{{\mathcal B}}
\newcommand{\caC}{{\mathcal C}}

\newcommand{\caF}{{\mathcal F}}

\newcommand{\caH}{{\mathcal H}}

\newcommand{\caJ}{{\mathcal J}}

\newcommand{\caL}{{\mathcal L}}

\newcommand{\caO}{{\mathcal O}}
\newcommand{\caP}{{\mathcal P}}

\newcommand{\caR}{{\mathcal R}}

\newcommand{\caT}{{\mathcal T}}
\newcommand{\caU}{{\mathcal U}}

\newcommand{\caW}{{\mathcal W}}

\newcommand{\bbC}{{\mathbb C}}

\newcommand{\bbE}{{\mathbb E}}

\newcommand{\bbN}{{\mathbb N}}

\newcommand{\bbP}{{\mathbb P}}

\newcommand{\bbR}{{\mathbb R}}

\newcommand{\opunit}{\text{1}\kern-0.22em\text{l}}

\newcommand{\e}{{\mathrm e}}
\renewcommand{\i}{{\mathrm i}}
\renewcommand{\d}{{\mathrm d}}
\newcommand{\sys}{{\mathrm S}}
\newcommand{\res}{{\mathrm R}}

\newcommand{\beq}{ \begin{equation} }
\newcommand{\eeq}{ \end{equation} }
\newcommand{\bet}{ \begin{theorem} }
\newcommand{\eet}{ \end{theorem} }
\renewcommand{\sp}{{\mathrm{sp}}}
\newcommand{\refl} {\mathrm{  Refl  }}
\newcommand{\systeemhamiltoniaan} {H_\sys}

\newcommand{\ra}{ \rightarrow }

\newtheorem{vetremark}[theorem]{\bf{Remark}}

\begin{document}
\begin{center}
\noindent{\large \bf  Steady state fluctuations of the dissipated
heat for a
quantum stochastic model} \\

\vspace{15pt}

{\bf Wojciech De Roeck}\footnote{Aspirant FWO, U.Antwerpen, email:
{\tt wojciech.deroeck@fys.kuleuven.be}} and {\bf Christian
Maes}\footnote{email: {\tt
christian.maes@fys.kuleuven.be}}\\
Instituut voor Theoretische Fysica\\ K.U.Leuven, Belgium.

\end{center}

\vspace{20pt} \footnotesize{ \noindent {\bf Abstract: }  We
introduce a quantum stochastic dynamics for heat conduction. A
multi-level subsystem is coupled to reservoirs at different
temperatures.  Energy quanta are detected in the reservoirs
allowing the study of steady state fluctuations of the entropy
dissipation. Our
main result states a symmetry in its large deviation rate
function.}

\vspace{5pt}

 \vspace{20pt} \normalsize

\section{Introduction}
Steady state statistical mechanics wants to construct and to
characterize the stationary distribution of a subsystem in contact
with several reservoirs.  By nature the required scenario is an
idealization as some essential specifications of the reservoirs
must be kept constant. For example, intensive quantities such as
temperature or (electro-)chemical potential of the different
reservoirs are defined and unchanged for an extensive amount of
time, ideally ad infinitum.  Reservoirs do not interact directly
with each other but only via the subsystem; they remain at their
same spatial location and can be identified at all times. That
does not mean that nothing happens to the reservoirs; flows of
energy or matter reach them and they are like sinks and sources of
currents that flow through the subsystem.  Concrete realizations
and models of steady states vary widely depending on the type of
substances and on the nature of the driving mechanism.

 An old and
standard problem takes the subsystem as a solid in contact at its
ends with two heat reservoirs and wants to investigate properties
of the energy flow. Beloved by many is a classical model
consisting of a chain or an array of coupled anharmonic
oscillators connected to thermal noises at the boundaries.  The
reservoirs are there effectively modeled by Langevin forces while
the bulk of the subsystem undergoes a Hamiltonian dynamics, see
e.g. \cite{reybelletthomas,eckmanpillet,maesnetocnyverscheure}.
Our model to be specified below is a quantum analogue of that
scenario in the sense that we also consider a combination of
Hamiltonian dynamics and Markovian thermal noises.

We imagine a chain of coupled two (or multi)-level systems. The
dynamics of the isolated subsystem is unitary with Hamiltonian
$H_\sys$. Quanta of energy $\omega$ are associated to the
elementary transitions between energy levels. Two physical
reservoirs at inverse temperatures $\beta_k, k=1,2$, are now
attached to the subsystem. The total dynamics is described by a
quantum stochastic differential equation through which we can
observe the number $N_{\omega,k}$ of quanta with energy $\omega$
that are piled up in the $k-$th reservoir. The total energy
$\mathbf{N} := H_\sys + \sum_{\omega,k} \omega
\mathbf{N}_{\omega,k}$ is conserved under the dynamics
(Proposition \ref{conserv}).  The change in the second term
corresponds to the flow of energy quanta in and out of the
reservoirs and specifies the dissipated heat. Our main result
consists in obtaining a symmetry in the fluctuations of that
dissipated heat that extends the so called steady state
fluctuation theorem for the entropy production to a quantum regime
(Proposition \ref{ft}).

 The quantum stochastic evolution that
defines the model is a particular dilation of a semigroup dynamics
that describes the weak coupling regime of our subsystem coupled
to quasi-free boson fields.  The dilation, a sort of quantum
Langevin equation, is much richer and enables the introduction of
a natural path space measure.  One should remember here that a
major conceptual difficulty in coming to terms with the notion of
a variable entropy production for quantum steady states is to
understand its path-dependence. One option is to interrupt the
unitary dynamics with collapses, see e.g. \cite{callensderoeck}.
Others have proposed an entropy production operator, avoiding the
problem of path-dependence. Our set-up follows a procedure that is
well-known in quantum optics with thermal noises formally
replacing photon detectors, see \cite{boutenguta,boutenkummerer}.
In the resulting picture we record each energy quantum that is
transferred between subsystem and reservoirs. It induces a
stochastic process on quanta transferrals and there remains no
problem to interpret the fluctuations of the entropy production.
From the mathematical point of view, the model can be analyzed via
standard probabilistic techniques.

\subsection{Related results} In the past decade, a lot of interest
has been going to the Gallavotti-Cohen fluctuation relation
\cite{gallavotticohen2,gallavotticohen95prl,evanscohen}, see
\cite{maesoriginanduse} for more recent references. In its
simplest form that relation states that the steady state
probability ($\Prob$) of observing a total entropy decrease
$w_T=-wT$ in a time $T$, is exponentially damped with respect to
the probability of observing an increase of $w_T$ as
\begin{equation}\label{entproform2}
\frac{\Prob(w_T=wT)}{\Prob(w_T=-wT)} \approx e^{+wT}
\end{equation}
at least for very large time spans $T$. The relation
(\ref{entproform2}) is known as the steady state fluctuation
theorem (SSFT) and states a symmetry in the fluctuations of the
entropy dissipation in a stationary nonequilibrium state. The
symmetry was first discovered in the context of dynamical systems
and was applied to the phase space contraction rate in strongly
chaotic dynamical systems, see
\cite{ruellesmooth,evanscohen,gallavotticohen2,gallavotticohen95prl}.
It was first further developed for stochastic dynamics in
\cite{kurchanstochastic,lebowitzspohn2,maes99}. In the present
paper, we deal with the SSFT for a quantum system. The steady
state condition must however be understood in a physical sense; it
is about heat conduction for fixed reservoirs in the long time
limit. The small system is treated in the steady state of an
approximate dynamics (the weak coupling limit) while the
reservoirs are kept at a fixed termperature.  Yet, mathematically,
we are not quite dealing with a steady or stationary state.  The
true total dynamics for system plus reservoirs is in fact much
more complicated.  We still speak about the steady state (and the
SSFT) also to contrast it with transient versions of the
fluctuation symmetry \eqref{entproform2}, see also
\cite{gallavotticohentwo}. Transient fluctuation theorems (TFT)
start typically from a change of variables at a finite time $t$,
reversing so to say the evolution, see \cite{sarmanevans}, and can
be obtained equally well for classical as for quantum systems.
That is not at all what we are doing here. We truly concentrate on
the stationary heat dissipation in the reservoirs, but from a
technical point of view, one could argue that our set-up is
actually a ``transient model of a steady state".

 The basic underlying mechanism and general unifying
principles connecting SSFT and TFT with statistical mechanical
entropy have been explained in
\cite{maesnetocny,maesoriginanduse}.

 Monnai and Tasaki \cite{monnaitasaki} have investigated
an exactly solvable harmonic system and found quantum corrections
to both SSFT and TFT. Matsui and Tasaki \cite{matsuitasaki} prove
a quantum TFT in a general $C^*$-algebraic setting. It is however
unclear what is the meaning of their entropy production operator.

A related quantum Jarzynski relation was studied in
\cite{deroeckmaes}.

 Besides the fluctuation theorem, we also
describe a new approach to the study of heat conduction in the
quantum weak coupling limit. In \cite{lebowitzspohn1}, Lebowitz
and Spohn studied the thermodynamics of the weak-coupling
generator. They identified the mean currents, and they proved a
Green-Kubo relation. At that time it was however not yet possible
to conclude that these expressions are the first non-zero
contributions to their counterparts at finite coupling $\lambda$.
That has recently been shown in a series of papers by Jak${\check
{\mathrm s}}$i\'{c} and Pillet
\cite{jaksicpillet1,jaksicpillet2,jaksicpillet3,jaksicpillet4},
who used spectral techniques to study the system at finite
coupling $\lambda$. It was also shown that the stationary state of
the weak coupling generator is the zeroth order contribution to
the system part of the so called NESS, the nonequilibrium steady
state. The current fluctuations we define in our model, agree with
the expressions of \cite{lebowitzspohn1} as far as the mean
currents and the Green-Kubo formula is concerned. Our entropy
production operator is however new; it differs for example from
the proposal of \cite{matsuitasaki}. The approach taken here also
differs from the more standard route that has been followed and
that was outlined by Ruelle in \cite{ruelleness}. Recently and
within that approach and context of heat conduction, new results
have been obtained in
\cite{jaksicpillet3,jaksicpillet4,abousalemfrohlich}. To us it
remains however very much unclear how to define and study in that
scenario a fluctuating entropy; in contrast, that is exactly one
of the things we can easily achieve via our approach but we remain
in the weak coupling limit.
\subsection{Basic strategies}
\subsubsection{Microscopic approach} \label{micro}
In general one would like to start from a microscopic quantum
dynamics.  The system is then represented by a finite-dimensional
Hilbert space $\mathcal{H}$ and system Hamiltonian $H_\sys$.  The
environment is made from thermal reservoirs, indexed by $k \in K$,
infinitely extended quantum mechanical systems, with formal
Hamiltonian,
\[
H_\res := \sum_{k \in K} H_{\res_k}
\]
The coupling between system and reservoirs is local and via some
bounded interaction term $\lambda H_{\sys-\res}$ so that the total
Hamiltonian takes the form
\[
H_\lambda :=  H_\sys \otimes 1 + 1 \otimes H_\res + \lambda
\sum_{k \in K} V_k \otimes R_k
\]
where we have already inserted a specific form for the coupling
$H_{\sys-\res}$ using self-adjoint reservoir operators $R_k$ and
$V_k$ acting on respectively $\caH_{\res_k}$ and $\mathcal{H}$.
On the same formal level, which can however easily be made
precise, the total quantum dynamics is then just
\[
U_t^\lambda := e^{-\i H_\lambda t}
\]
We will not follow the beautiful spectral or scattering approach
that has recently been exploited for that nonequilibrium problem.
We refer the reader to the specialized references such as
\cite{jaksicpillet3,ruelleness} and we only outline the main
steps, totally ignoring essential
assumptions and technicalities:\\
One starts the dynamics from an initial state
\[
\mu := \rho_\sys \otimes \rho_{\res_1} \otimes \ldots \otimes
\rho_{\res_{|K|}}
\]
where $\rho_\sys$ stands for an initial state in the system and
the $\rho_{\res_k}$ are equilibrium KMS states at inverse
temperature $\beta_k$ for the $k$-th reservoir. The quantum
dynamics takes that initial state to the new (now coupled) state
$\mu_t$ at time $t>0$.  The NESS is obtained via an ergodic
average \beq \label{def: ness} \mu_{\mathrm{NESS}} :=
\lim_{T\uparrow+\infty} \frac 1{T} \int_0^Tdt\, \mu_t \eeq One of
the first questions (and partially solved elsewhere, see e.g.
\cite{jaksicpillet3,ruelleness,aschbacherspohn}) is then to derive
the natural conditions under which the mean entropy production
rate
\[
\dot{S} := \i \sum_{k=1}^m\beta_k \,\mu_{\mathrm{NESS}}(
[H_\lambda,H_{\res_k}])
\]
is strictly positive.  While that mean entropy production
certainly coincides with conventional wisdom, we do not however
believe that the operator
\[
\i [H_\lambda,H_{\res_k}]
\]
or equivalent expressions, is the physically correct candidate for
the study of current fluctuations which would obey the SSFT.  That
is not even the case for the simplest (classical) stochastic
dynamics; one needs to go to path space and study current
fluctuations in terms of (fluctuating) trajectories.

\subsubsection{Weak coupling approach}\label{weakapproach}

Starting from the microscopic dynamics above, we can of course
always look at the reduced dynamics $\Lambda^\lambda_t$ on the
system
\[
\Lambda_t^\lambda\rho_\sys = \Tr_R \left[U_t^\lambda(\rho_\sys
\otimes \rho_{\res_1} \otimes \ldots \otimes \rho_{\res_{|K|}})
U_{-t}^\lambda\right] \]
 for a density matrix $\rho_\sys$ on the system.  Obviously, the
 microscopic evolution couples the system with the environment and
 the product form of the state will in general not be preserved.
 One can however attempt a Boltzmann-type Ansatz or projection
 technique to enforce a repeated randomization.  That can be made
 rigorous in the so called weak coupling limit.  For that, one
 needs the interaction picture and one keeps $\lambda^2 t =\tau$
 fixed.  That is the Van Hove-Davies-limit
 \cite{vanhove,davies1}
 \[
\lim_{\lambda\rightarrow 0} \Lambda_{-t}^0\,\Lambda_t^\lambda
\rho_\sys := e^{\tau \caL^*} \rho_\sys
\]
where $\caL^*$ is a linear operator acting on density matrices for
the system.  The generator will be written out more explicitly in
Section \ref{weak} but its dual $\mathcal{L}$ acting on
$\mathcal{B}(\mathcal{H})$ is of the form, see
\eqref{decomposition gen},
\[
 \mathcal{L}(\cdot)= \i [H_{f},\cdot]+\sum_{k \in K} \mathcal{L}_k(\cdot)
\]
where the $\mathcal{L}_k$ can be identified with the contribution
to the dissipation from the $k$'th reservoir. $H_f$ is an effective, renormalized Hamiltonian depending on details of the reservoirs and the coupling.\\
From now on, we write $\rho$ for the (assumed) unique invariant
state (see also Remark \ref{rem: ergodicity}):
\[
e^{\tau \caL^*}\rho =\rho,\quad \tau\geq 0
\]
Again one can study here the mean entropy production, as for
example done in \cite{lebowitzspohn1} and argue that
\[
\mbox{Tr}[\rho\,\mathcal{L}_k H_\sys ]
\]
represents the stationary heat flow into the $k$'th reservoir, at
least in the weak coupling regime.  Nothing tells us here however
about the physical fluctuations in the heat current for which
higher moments should be considered.  In fact, the reservoirs are
no longer visible as the weak coupling dynamics is really a jump
process on the energy levels of the system Hamiltonian, see
further in Section \ref{clasweak}.  The heat flow and the energy
changes in  the individual reservoirs cannot be reconstructed from
the changes in the system.  The present paper uses a new idea
for the study of the fluctuations of the heat dissipation in a
reservoir.

\subsubsection{Dilation} \label{sec: dilation}
While the weak coupling dynamics is very useful for problems of
thermal relaxation (one reservoir) and for identifying the
conditions of microscopic reversibility (detailed balance)
characterizing an equilibrium dynamics, not sufficient information
is left in the weak coupling limit to identify the variable heat
dissipated in the various reservoirs.  Heat is path-dependent and
we need at least a notion of energy-trajectories.\footnote{At
least, if one has a stochastic or effective description of the
system dynamics, as is the case in the weak-coupling limit. We do
{\bf not} claim at all that the trajectory-picture is
microscopically fundamental.} The good news is that we can obtain
such a representation at the same time as we obtain a particular
dilation of the weak coupling dynamics. The representation is
basically achieved via an unraveling of the weak coupling
generator $\mathcal{L}$ and the corresponding Dyson expansion of
the semigroup dynamics.  That will be explained in Section
\ref{unravel}.

 There are many possible dilations of a quantum
dissipation.  It turns out that there is a dilation whose
restriction to the system coincides with the Dyson representation
in terms of energy-trajectories of the weak coupling dynamics.
That dilation is well studied and goes under the name of a quantum
stochastic dynamics.  The associated quantum stochastic calculus
was invented by Hudson and Parathasaraty,
\cite{hudsonparathasaraty}. It has been extensively employed for
the purpose of quantum counting processes, see e.g.
\cite{boutenguta,boutenkummerer}. Various representations and
simplifications have been added, such as in \cite{alickifannes}
where a (classical) Brownian motion extends the quantum
dissipation. Unravelings of generators have been first employed in
quantum optics in \cite{srinivasdavies}, they are further
discussed in \cite{carmichael}.
\subsubsection{Results}
We prove a symmetry in the large deviation generating function of
the dissipated heat (Proposition \ref{thm: ft}). This function is
analytic and this implies the large deviation principle. The
symmetry is recognized as the fluctuation theorem for the entropy
production. The precise form of the fluctuation theorem depends on
whether the model has been derived from a reversible or an
irreversible (e.g.\ because of the presence of magnetic fields)
dynamics. This point was clarified in \cite{maesirreversible}. By
a theorem of Bryc \cite{bryc}, analyticity of the generating
function implies the  central limit theorem for the currents. We
do not stress this point but it is implicitly used in deriving a
Green-Kubo relation and Onsager reciprocity (Proposition \ref{thm:
green kubo}), or modifications of these, again depending on the
reversibility of the original model. In all cases, the fluctuation
symmetry helps to establish strict positivity of the entropy
production (Proposition \ref{thm: strict positivity}). Let us
stress that our main result, Proposition \ref{thm: ft}, depends on
an interpretation, as described above under Section \ref{sec:
dilation}. However, the consequences of our main result,
Propositions  \ref{thm: strict positivity}  and \ref{thm: green
kubo} do not depend on this interpretation. This will be further
discussed in Section \ref{sec: discussion}.

\subsubsection{Comparison with earlier results}
Technically, our fluctuation theorem is very close to the results
obtained in \cite{kurchanstochastic} or \cite{lebowitzspohn2}. The
Green-Kubo relations and Onsager reciprocity have been established
recently in e.g. \cite{jaksicogata} for the spin-fermion model. In
the weak-coupling limit they were discussed already in
\cite{lebowitzspohn1}, however there the authors did not
distinguish between reversible and irreversible models (this is
commented upon in Remark \ref{rem: degenerate time-reversal}). The
strict positivity in the weak-coupling limit was proven in
\cite{jaksicpillet3}\footnote{ Besides, from \cite{jaksicpillet3},
it follows that the strict positivity remains true at small
nonzero coupling, without taking the weak coupling limit.} (for
the spin-fermion model) and in \cite{aschbacherspohn} (under
general conditions).
 Our theorem on strict positivity is however
slightly more general: Assuming the existence of a unique,
faithful stationary state, we formulate a necessary and sufficient
condition for strict positivity.

\subsection{Outline of the paper}
In Section 2, we introduce the quantum stochastic model and state
the result. In Section 3 follows a discussion where the main
points and novelties are emphasized. Proofs are postponed to
Section 4.

\section{The Model}
\subsection{Weak Coupling}\label{weak}
We briefly introduce here the weak coupling dynamics without
speaking about its derivation, which is not relevant for the
discussion here. Some of that was briefly addressed in Sections
\ref{micro}-\ref{weakapproach} and it is covered in detail in
\cite{davies1} and \cite{lebowitzspohn1}.

 Let $\mathcal{H}$ be a
finite-dimensional Hilbert space assigned to a small subsystem,
called system in what follows. Let $H_\sys$ be a self-adjoint
Hamiltonian on $\mathcal{H}$. Introduce the set of Bohr
frequencies
\begin{equation}\label{setomega}
F := \{\omega \in \mathbb{R} \, | \, \exists \, e,e' \in \sp
H_\sys : \omega=e-e' \}
\end{equation}
Remark that $F$ is the set of eigenvalues of the derivation $-\i
[H_\sys,\cdot]$. We label by $k \in K$ (a finite number of)
different heat reservoirs at inverse temperatures $\beta_k
<\infty$. To each reservoir $k$ is assigned a self-adjoint
operator $V_k \in \caB(\caH)$  and for each $k \in K, \omega \in
F$, we put
\begin{equation} \label{newv}
V_{\omega,k}= \sum_{\footnotesize{\left.\begin{array}{cc} e,e' \in
\sp H_\sys \\ \om=e-e'\end{array}\right.} } 1_e (\systeemhamiltoniaan)
V_k 1_{e'}(\systeemhamiltoniaan)
\end{equation}
where $1_e (H_\sys)$ for $e \in \sp(H_\sys)$ is the spectral
projection on $e$ associated to  $H_\sys$.


Fix for $k \in K$, nonnegative functions $\eta_k \in L^1(\bbR)$
and assume them to be  H\"older continuous in $F\subset \bbR$ and
satisfying the condition \beq \label{KMS condition} \eta_k (x)
=\e^{-\beta_k x} \eta_k (-x) \geq 0, \quad x \in \bbR \eeq which
is related to the KMS equilibrium conditions in the reservoir $k
\in K$, see further under remark \ref{rem: correlation functions}.
Write also for $\om \in F, k \in K$ \beq s_k(\omega)=\lim_{\ep
\downarrow 0} \int_{\bbR \setminus [\om-\ep,\om+\ep]}
\frac{\eta_k(x)}{\om-x} \eeq which is well defined by the
assumption of H\"older continuity for $\eta_{k \in K}$. From now
on, we simply write the indices ${\om,k}$ for ${\om \in F,k\in
K}$.
We consider the self-adjoint Hamiltonian
\begin{equation}
H_{f}:= \sum_{\om,k} s_k(\om)  V^*_{\om,k} V_{\om,k}
\end{equation}
satisfying by construction
\begin{equation}
[H_{f},H_\sys]=0
\end{equation}
We work with the following generator $\mathcal{L}$ on
$\mathcal{B}(\mathcal{H})$
\begin{equation} \label{gen2}
\mathcal{L}(\cdot)= \i [H_{f},\cdot]+\sum_{\omega,k}
\eta_k(\omega) \big(
 V_{\omega,k}^{*}\cdot
V_{\omega,k}-\frac{1}{2}\{V_{\omega,k}^{*}V_{\omega,k},\cdot\}
\big)
\end{equation}
 Putting $\caT(\caH) \subset \caB(\caH)$ the set of all
density matrices on $\caH$, i.e: \beq \mu \in \caT(\caH)
\Leftrightarrow \Tr \left[ \mu \right]=1, \qquad \mu \geq 0 \eeq one introduces
the dual generator $\caL^*$ on $\caT(\caH)$, defined through \beq
\Tr \left[ A \caL^*\mu  \right]  = \Tr \left[\mu \caL A \right] ,  \qquad A \in \caB(\caH), \mu \in \caT(\caH) \eeq
By grouping all terms with the same $k$ in \eqref{gen2}, we can
also write
\begin{equation} \label{decomposition gen}
 \mathcal{L}(\cdot)=-\i [H_{f},\cdot]+\sum_{k \in K} \mathcal{L}_k(\cdot)
\end{equation}
Both $\caL$ and $\caL_{k \in K}$ are of the Lindblad form
\cite{lindblad} and hence they generate completely positive
semigroups $\e^{t \caL}$ and $\e^{t \caL_k}$. A $\rho \in
\caT(\caH)$ is a \emph{stationary state} for the semigroup $\e^{t
\caL}$ iff \beq \caL^* \rho=0 \quad \textrm{or, equivalently,}
\quad \e^{t \caL^*} \rho=\rho \eeq
%
%
%
We  fix an anti-unitary operator $T$ on $\caH$, which has to be
thought of as playing the role of time reversal. Let \beq
\label{def: time-reversed model} H^{\theta}_S := T H_\sys T ,
\qquad V^{\theta}_k := T V_k T
 \eeq
That defines a new model, satisfying all necessary requirements.
This model can be thought of as the time-reversal of the original
one.

We will need the following assumptions:

\noindent {\bf Assumption A1}\\
\textit{We ask triviality of the commutant \beq \label{cond:
trivial commutant} \{ \eta^{1/2}_k(\om)V_{\om,k } \, | \, k \in K,
\om \in F \} '= \bbC 1 \eeq where for $\caA \subset \caB(\caH)$,
\beq B \in  \caA '  \Leftrightarrow \forall A \in \caA : [A,B]=0
\eeq
That ensures the existence of a unique stationary state, as stated in Remark \ref{rem: ergodicity}}\\

\noindent {\bf Assumption A2}\\
\textit{We ask that the system can
complete a closed cycle in which the entropy production is
nonzero. More precisely, there are
sequences $\om_1,\ldots, \om_n  $ in $F$ and  $k_1,\ldots, k_n$ in
$K$ such that
\begin{enumerate}
\item{ \beq\label{cond: cycle nonzero ent} \sum_{i=1}^n  \beta_{k_i} \om_i \neq 0  \eeq }
\item{There is a one-dimensional projection $P \in \caB(\caH)$ such that
\beq \label{cond: closed cycle} \Tr \left[ P  V_{\om_n,k_n}
\ldots V_{\om_2,k_2}  V_{\om_1,k_1} P \right]
\neq 0 \eeq }
\end{enumerate}
}

\noindent{\bf Assumption A3}\\
\textit{This assumption expresses that our model is time-reversal
invariant. It will be used in deriving the full fluctuation
theorem, the Green-Kubo relations and Onsager reciprocity.
 \beq   H^{\theta}_S =H_\sys,    \quad   \forall k \in K: \, V^{\theta}_{k}  =V_{k}      \eeq
 }

\begin{vetremark}\label{rem: ergodicity}
If Assumption {\bf A1} holds, then, by a theorem of Frigerio
(Theorem 3.2 in \cite{frigerio}) and the fact that $\be_{k \in K}
< \infty$, the semigroup $\e^{t \caL}$ has a unique stationary
state $\rho$. This state is faithful, i.e.\ for all nonzero
projections $P \neq 0 \in \caB(\caH)$: \beq \Tr \left[ \rho P
\right]>0 \eeq Assumption {\bf A1} is actually a necessary
condition for the existence of a unique stationary state.
\end{vetremark}

\begin{vetremark}
Assumption {\bf A2} comprises the intuitive assumption that the
system does not break up in independent subsystems which are
coupled separately to the reservoirs. If that would be the case,
then most of our results still hold but they become trivial. For
example, the rate function $e$ from Proposition \ref{thm: ft}
satisfies $\forall \ka \in \bbC: \, e(\ka) = 0  $.
\end{vetremark}

\begin{vetremark}\label{rem: canonical}
If for all $k \in K$, $\be_k=\be$ for some $\be$, then \beq
\rho_{\beta} :=\exp(-\beta H_\sys)/ \Tr \left[ \exp(-\beta H_\sys)
\right]\eeq is a stationary state for $\e^{t \caL}$, as follows
from the condition \eqref{KMS condition} and the explicit form
\eqref{gen2}.
\end{vetremark}

\begin{vetremark} \label{blockdiagonal}
If {\bf A1} holds (assuring the uniqueness of the stationary
state), then one easily checks  \beq \label{decoherence} \forall A
\in \caB(\caH), \forall e \neq e' \in \sp H_\sys : \, \lim_{t
\uparrow +\infty} \e^{t \caL} \big(1_e(H_\sys) A 1_{e'}(H_\sys)
\big) =0 \eeq which is usually called ``decoherence". As a
consequence of \eqref{decoherence}, the stationary state $\rho \in
\caT(\caH)$ of $\e^{t \caL}$ satisfies,
\begin{equation}\label{rhodiag}
\sum_{ e \in \sp H_\sys} 1_e (H_\sys) \, \rho \, 1_e (H_\sys)=\rho
\end{equation}

\end{vetremark}

\begin{vetremark} \label{rem: correlation functions}
If one would derive the model from a microscopic setup, then we
can be more specific. Let $\caH_{R_k}$ be the Hilbert space of the
$k$'th reservoir and $\rho_k$  a thermal equilibrium state at
$\be_k$ on (a subalgebra of) $\caB(\caH_{R_k})$. Assume the
coupling is given by \beq \label{full coupling}\sum_{k \in K} V_k
\otimes R_k,
 \qquad R_k=R^*_k \in \caB(\caH_{R_k}) \eeq
Then the functions $\eta_k$ are fourier transforms of the
autocorrelation function of $R_k$ and the KMS conditions imply
\eqref{KMS condition}. All this is discussed at length in
\cite{lebowitzspohn2}.

 The restriction
to couplings of the form \eqref{full coupling}, where each term is
self-adjoint by itself, is not necessary. Besides, one can also
have multiple couplings per reservoir. Since this complicates our
notation without introducing any novelty, we adhere to the simple
form \eqref{full coupling}.
 \end{vetremark}


\begin{vetremark} \label{rem: degenerate time-reversal}
If $H_\sys$ is nondegenerate, one can choose $T$ as follows: Let
$\psi_e, \, e \in \sp H_\sys$ be a complete set of eigenvectors
for $H_\sys$ and put \beq T \big(\sum_{e \in \sp H_\sys} c_e
\psi_e \big) =  \sum_{e \in \sp H_\sys} \bar{c}_e  \psi_e , \qquad
c_{e \in \sp H_\sys} \in \bbC \eeq Although this does not
necessarily imply Assumption {\bf A3}, it does imply \beq
  H^{\theta}_S =H_\sys ,  \quad  V_{\om,k} A  V^*_{\om,k} =  V^{\theta}_{\om,k} A  (V^{\theta}_{\om,k})^*  \qquad A \in \caB(\caH)
\eeq which, as one can check from the proofs, can replace {\bf A3}
for all purposes of this paper. Hence, a nondegenerate model is
automatically time-reversal invariant. This explains why in
\cite{lebowitzspohn1} the Green-Kubo relations were derived for
nondegenerate Hamiltonians without speaking about microscopic
time-reversal. It also explains why time-reversal does not appear
naturally in the framework of classical Markov jump processes.
\end{vetremark}
\subsection{Unraveling the generator}\label{unravel}
We associate to that semigroup dynamics, generated by
\eqref{gen2}, a pathspace measure by a procedure which is known as
``unraveling the generator". Basically, we will introduce
$|F|\times |K|$ Poissonian clocks, one for each reservoir and each
Bohr frequency. Whenever clock $(\om,k)$ ticks, our system will
make a transition with Bohr frequency $\om$, induced by reservoir
$k$. This will be our `a priori' measure $\d \si$ (see further).
If $H_\sys$ is nondegenerate, then it is very easy to upgrade $\d
\si$ to the appropriate pathspace measure: one multiplies $\d \si$
with a certain factor for each jump and with factors for the
waiting times, obtaining something of the form \beq  \d
\bbP_{\rho_0} (\si) =   \e^{-(t-t_n) r_{n+1}}    c_{n} \ldots
c_2\e^{-(t_2-t_1) r_1} c_1 \e^{-t_1 r_1} \d \si  \eeq for some
positive numbers $c_1,\ldots,c_n$ and $r_1,\ldots,r_{n+1}$ and
initial state $\rho_0$.

When $H_\sys$ is degenerate, one has to do things more carefully,
leading to the expression \eqref{maat} in Lemma \ref{prob}. The
technical difference between degenerate and non-degenerate
$H_\sys$ is further discussed in Section \ref{sec: discussion}.
\\

\subsubsection{Preliminaries}
Put \beq \Omega^1_{t} := \{ \si \subset [0,t] \big | |\si| <
\infty \}, \qquad \Omega^1 :=\{ \si \subset \bbR^+ \big | |\si| <
\infty \} \eeq where $|\si|$ is the cardinality of the set $\si
\subset \bbR$. Let $(\Omega^1_t)_{\omega,k}, \Omega^1_{\omega,k} $
stand for identical copies of $\Omega^1_t, \Omega^1$ and put \beq
\label{product Omegas}  \Omega :=
\mathop{\times}\limits_{\omega,k} \Omega^1_{\omega,k},
   \qquad  \Omega_t :=  \mathop{\times}\limits_{\omega,k}
(\Omega^1_t)_{\omega,k}    \eeq $\Omega$ and $\Om_t$ are called
Guichardet spaces, see \cite{guichardet}. An element $\sigma \in
\Omega$ looks like
\begin{equation}\label{ordered}
\sigma=(\omega_1,k_1,t_1;\ldots;\omega_n,k_n,t_n)  \textrm{  with
}  0<t_1 < t_2 < \ldots < t_n <+\infty
\end{equation}

\noindent Alternatively, and corresponding to the product in
\eqref{product Omegas}: \beq \label{product omegas} \si =(
\si_{\omega,k})_{\om,k} \textrm{ with } \si_{\omega,k} \in
\Omega^1_{\om,k}, \qquad |\si|:= \sum_{\om,k} |\si_{\om,k}| \eeq

\noindent We define integretation on $\Om_t$ and $\Om$, by putting
for any sequence of  functions $g=(g_n)_{n \in \bbN}$ with $g_n$ a
measurable function on $ F^n \times K^n \times (\bbR^+)^n $ for
all $n \in \bbN$, \beq \label{def: integration}\int_{\Om_t} \d \si
g(\si) :=\sum_{n=0}^{\infty}\sum_{ \footnotesize{ \left.
\begin{array}{l} k_1,\ldots,k_n \in K^n
\\
\om_1,\ldots,\om_n \in F^n
\end{array} \right.} }  \int_{\triangle_t^n} \d t_1 \ldots \d
t_n   g_n \big( (\om_1,k_1,t_1;\ldots;\om_n,k_n, t_n ) \big)
 \eeq
 where $\triangle_t^n \subset \bbR^n$ is the simplex
 \beq
 (t_1,\ldots,t_n) \in \triangle_t^n \Leftrightarrow 0<t_1 < \ldots <
 t_n <t
 \eeq
The equality \eqref{def: integration} defines the symbol ``$\d
\si$" and the notion of measurable sets in $\Om_t$ or $\Om$ (for
the latter, take $t=\infty$ in the above definitions).

For future use, we introduce 'number functions' $n^t_{\omega,k}$,
defined as
\begin{equation}\label{numbers}
 n^t_{\omega,k}(\sigma):= |\sigma_{\om,k} \cap [0,t]| , \qquad n_{\omega,k}(\sigma):= |\sigma_{\om,k} |
\end{equation}
and the abbreviations $\si \cup \tau$ and $\tau \setminus \si$ for
elements of $\Om$, defined by
\begin{equation}\label{cups in Om}
  (\si \cup \tau)_{\om,k}:= \si_{\om,k} \cup \tau_{\om,k} ,   \qquad   \si= \xi \setminus \tau \Leftrightarrow \si \cup \tau =\xi
\end{equation}
If $\si \in \Om_s$ and $\tau \in \Om_u$, we also need $ \si \tau \in \Om_{s+u}$, defined by
\begin{equation}\label{glue in Om}
  ( \si \tau)_{\om,k} := \si_{\om,k}
\cup (s+\tau_{\om,k})     \quad \textrm{where  } q \in
s+\tau_{\om,k} \Leftrightarrow q-s \in \tau_{\om,k}
\end{equation}
Remark that a function $g$ on $\Om_s$ is naturally made into a
function on $\Om_{s+u}$ by, using the notation \eqref{glue in Om},
\beq \label{extending function} g(\si \tau)  :=    g(\si)
,\qquad     \si \in \Om_s, \tau \in \Om_u \eeq

\subsubsection{Constructing a pathspace measure}

Write the weak coupling generator \eqref{gen2} as
\beq\label{splitting} \caL=\caL_0+\sum_{\omega,k} \caJ_{\omega,k}
\eeq with
\begin{equation}\label{j}
\caJ_{\omega,k}(\cdot):=  \eta_k(\om)  V^*_{\omega,k} \cdot V_{\omega,k}
\end{equation}
 and
\[
\caL_0(\cdot):=\i [H_{f},\cdot] -\frac{1}{2}\sum_{\omega,k}
\eta_k(\omega) \{V_{\omega,k}^{*}V_{\omega,k},\cdot\}
\]
Consider $\mathcal{W}_t(\sigma): \mathcal{B}(\mathcal{H})
\rightarrow \mathcal{B}(\mathcal{H})$ as the completely positive
map depending on $\sigma \in \Omega$
\begin{equation} \label{explicitmap}
\mathcal{W}_t(\sigma):=I_{\Om_t}(\sigma) e ^{t_1\mathcal{L}_0}
\mathcal{J}_{\omega_1,k_1}e^{(t_2-t_1)\mathcal{L}_0}\ldots
e^{(t_{|\si|}-t_{|\si|-1})\mathcal{L}_0}\mathcal{J}_{\omega_{|\si|},k_{|\si|}}
e^{(t-t_{|\si|})\mathcal{L}_0}
\end{equation}
with $I_{\Om_t}$ the indicator function of $\Omega_{t} \subset
\Omega$ and with the indices $(\omega_i,k_i), i=1,\ldots,{|\si|}$
referring to the representation (\ref{ordered}) of $\sigma$. To
verify the complete positivity of \eqref{explicitmap}, rewrite
$\caL_0$ as
\beq \label{def: K} \caL_0 (\cdot) = S \cdot + \cdot S^* , \qquad
S= \i H_f -\frac{1}{2}\sum_{\om,k} \eta_k(\om)
V^*_{\om,k}V_{\om,k} \eeq which yields, \beq \label{expressing L0
in K} \e^{t\caL_0} (\cdot) = \e^{tS} \cdot (\e^{t S})^* \eeq
Complete positivity of \eqref{j} is obvious from its definition.
 The Dyson expansion of $\e^{t \caL}$, corresponding to the
splitting \eqref{splitting}, reads
\begin{equation}\label{def: dyson}
e^{t \caL}=\int_{\Omega_t} \d \sigma  \mathcal{W}_t(\sigma)
\end{equation}
That expression induces a `path space measure', or a notion of
`quantum trajectories' on $\Omega$.

\begin{lemma} \label{prob}  Choose $\mu \in \caT(\caH)$.
Let $E \subset \Om_t$ be measurable and define
\begin{equation}\label{maat}
\mathbb{P}_{\mu,t}(E) := \int_{E} \d \sigma \Tr \left[ \mu
\mathcal{W}_t(\sigma)1 \right]
\end{equation}
Then $(\mathbb{P}_{\mu,t})_{t \in \bbR^+}$ are a consistent family
of probability measures on $(\Omega_t)_{t \in \bbR^+}$, i.e.\ for
a measurable function $g$ on $\Om_t$,
 \beq \int_{\Om_t} \d\mathbb{P}_{\mu,t}(\si) g(\si)=
\int_{\Om_s} \d\mathbb{P}_{\mu,s}(\si) g(\si) ,\qquad     s  \geq
t
    \eeq
    where $g$ is extended to $\Om_s$ as in \eqref{extending function}.
\end{lemma}
\noindent Thus we obtain a new
 probability measure $\mathbb{P}_{\mu}$
on $\Omega$ by the Kolmogorov extension theorem, for $t
>0$ and a function $g$ on  $\Om_t$, \beq  \int_{\Om} \d\mathbb{P}_{\mu}(\si)  g(\si)
= \int_{\Om_t} \d\mathbb{P}_{\mu,t}(\si) g(\si) \eeq where we used
again the extension as in \eqref{extending function}. The
expectation with respect to these measures is denoted \beq
\mathbb{E}_{\mu,t} \textrm{ on } \Om_t, \qquad \mathbb{E}_{\mu}
\textrm{ on } \Om\eeq These probability measures are often called
`quantum counting processes', see
\cite{boutenguta,boutenkummerer}.
\subsection{Results}

We define the integrated entropy current $w^t$ up to time $t$ as a
function on $\Omega$:
\begin{equation}\label{entproclas}
w^t(\sigma) =-\sum_{\omega,k} \beta_{k} \omega
n^t_{\omega,k}(\sigma)
\end{equation}
with $n^t_{\omega,k}$ as in (\ref{numbers}). In what follows, we denote by $\rho$ the
stationary state for $\e^{t \caL}$, which is unique by Assumption {\bf A1}.  For $\ka \in \bbC$,
we write
\begin{equation}
e(\kappa) := \lim_{t \uparrow +\infty} \frac{1}{t}\log
\mathbb{E}_{\rho} [e^{-\kappa w^t}]
\end{equation}
if it exists. Then, $e(\ka)$ of course depends on all model
parameters, i.e.\ on $H_\sys, V_{k}, \eta_{k}$. We introduce
$e^{\theta}(\ka)$ which is derived from the model with new
parameters $ H^{\theta}_S , V^{\theta}_{k},
\eta^{\theta}_k=\eta_k$, see \eqref{def: time-reversed model}.

Now we can already formulate the main result of the paper:
\begin{proposition}\label{thm: ft} {\bf Fluctuation Theorem}\\ Assume {\bf A1}. Let $w^t$ be defined by
(\ref{entproclas}).
There is an open set $\caU \in \bbC$ containing the real line,
$\bbR \subset \caU$, such that for all $\ka \in \caU$, the limit
\begin{equation}\label{rate function clas}
e(\kappa) := \lim_{t \uparrow +\infty} \frac{1}{t}\log
\mathbb{E}_{\rho} [e^{-\kappa w^t}]
\end{equation}
exists and the function $\ka \to e(\kappa)$ is analytic on $\caU$.
Moreover,
\begin{equation} \label{symmetry2}
e(\kappa)=e^{\theta}(1-\kappa)
\end{equation}
If also {\bf A3} holds, then $e(\kappa)=e^{\theta}(\kappa)$ and
\begin{equation} \label{symmetry}
e(\kappa)=e(1-\kappa)
\end{equation}

\end{proposition}

\noindent We list some consequences of the fluctuation relations
\eqref{symmetry2} and \eqref{symmetry}.

\begin{proposition}\label{thm: strict positivity}{\bf Strict Positivity of the Entropy Production}\\
Assume {\bf A1}, then
\begin{equation} \label{nonneg}
\textrm{ {\bf A2} holds }  \Leftrightarrow    \lim_{t \uparrow
+\infty} \frac{1}{t}\mathbb{E}_{\rho} [w^t]
> 0
\end{equation}
\end{proposition}

\noindent For the next proposition we introduce energy functions
$n^t_{k}$ on $\Om$; \beq \label{energy functions} n^t_{k}:=
-\sum_{\om \in F} \om n^t_{\om,k}\eeq


\begin{proposition}\label{thm: green kubo}{\bf Green-Kubo Relations}\\
Assume {\bf A1} and fix some $\be > 0$. Let for $k,k' \in K$:
\begin{equation}
L_{k,k'}(\be) := \frac{\partial}{\partial \beta_{k'}}\lim_{t
\uparrow +\infty} \frac{1}{t} \mathbb{E}_{\rho}[n^t_k ]
\Big|_{\be_{1}=\ldots=\be_{|K|}=\be}
\end{equation}
and similarly the time-reversed coefficient $L^{\theta}_{k,k'}$,
obtained by starting with $H_\sys^{\theta}$ and $V_k^{\theta}$.
Then,
\begin{equation} \label{eq: modified green kubo}
L_{k,k'}(\be)+ L^{\theta}_{k,k'}(\be)=  \beta \lim_{t \uparrow
+\infty}  \frac{1}{t} \mathbb{E}_{\rho} [n^t_{k} n^t_{k'}]
\end{equation}
If also {\bf A3} holds, then
\begin{equation} \label{eq: green kubo}
L_{k,k'} = \frac{1}{2} \beta \lim_{t \uparrow +\infty}
\frac{1}{t} \mathbb{E}_{\rho} [n^t_{k} n^t_{k'}]
\end{equation}
with Onsager reciprocity
\begin{equation} \label{eq: onsager reciprocity}
L_{k,k'} = L_{k',k}
\end{equation}
\end{proposition}

\subsection{The quantum model: a dilation of the semigroup $\e^{t \caL}$}

\subsubsection{Heuristics}
In the next section we construct a unitary evolution, which is our
basic quantum model. This type of unitary evolutions is generally
known as solutions of quantum stochastic differential equations,
introduced in \cite{hudsonparathasaraty}.\\  For the readers who
are familiar with stochastic calculus, we briefly state how our
evolution would look in traditional notation. Recommended
references are \cite{parthasarathybook,attalreview} for quantum
stochastic calculus and \cite{derzinski1} for the formalism of
second quantization.

 For all $\omega \in F$ and $k
\in K$, let $\left( L^2(\bbR^+)\right)_{\omega,k}$ be a copy of
$L^2(\bbR^+)$. We consider the bosonic Fock space ($\Gamma_s$
denotes symmetrized second quantization)
\begin{equation}\label{fockspace}
\mathcal{R} = \Gamma_s \left [ \mathop{\oplus}\limits_{\omega,k}
\left( L^2(\bbR^+)\right)_{\omega,k}  \right]=
\mathop{\otimes}\limits_{\omega,k} \Gamma_s \left [ \left(
L^2(\bbR^+)\right)_{\omega,k} \right]
\end{equation}
and think of $\d\mathbf{A}^{*}_{\omega,k,t}$ with $t \in \bbR^+$
as the creation operator on $\Gamma_s \left [ \left(
L^2(\bbR^+)\right)_{\omega,k} \right]$ creating the
``wavefunction" $\chi_{[t,t+ \d t]}$ (the indicator function of
the interval $[t,t+\d t]$). We now write a Quantum Stochastic
Differential Equation (QSDE) on $\mathcal{B}(\mathcal{H} \otimes
\mathcal{R} )$:
\begin{eqnarray}\label{dilation1}
\d\mathbf{U}_t &=& \sum_{\omega,k} \eta^{1/2}_k(\om) \left(
V_{\omega,k} \d\mathbf{A}^{*}_{\omega,k,t}-V^{*}_{\omega,k}
\d\mathbf{A}_{\omega,k,t}\right)\mathbf{U}_t\\
&-&\left(\i H_{f} \d t-\frac{1}{2}\sum_{\omega,k} \eta_k(\om)
V^{*}_{\omega,k}V_{\omega,k} \d t \right) \mathbf{U}_t, \quad
\mathbf{U}_0=1\otimes \mathbf{1} \nonumber
\end{eqnarray}

Of course, the intuitive definitions given here, do not suffice to
give meaning to this expression. We content ourselves with stating
that \eqref{dilation1} defines a unitary evolution $\mathbf{U}_t$,
which we will now rigourously construct by using Maassen's
approach of integral kernels \cite{maassen1}.
%

%
%
\subsubsection{Construction of the unitary evolution $\mathbf{U}_t$}
Recall the Guichardet spaces $\Om_t$ and $\Om$, introduced in
section  \ref{unravel} and define for $(\si,\tau) \in \Om \times \Om $ the
ordered sequence of times $ (t_1,\ldots ,t_n)$ as \beq  \{t_1,\ldots
, t_n \} = \cup_{\omega,k} (\si_{\omega,k} \cup \tau_{\omega,k} )
\textrm{ and } 0< t_1< \ldots   <t_{n}   \quad
n=|\si|+|\tau| \eeq

We define the integral kernel $u_t: \Omega \times \Om
\hookrightarrow \caB(\caH)$:
\begin{equation}\label{kernel}
u_t(\sigma,\tau)= I_{\Omega_t \times
\Omega_t}(\sigma,\tau)\e^{(t-t_n ) K}
Z_{n}\e^{(t_n-t_{n-1})K}Z_{n-1} \ldots Z_2 \e^{(t_2-t_1) K}Z_1
\e^{t_1 K}
\end{equation}
with $I_{\Omega_t \times \Omega_t}$ the indicator function of
$\Omega_t \times \Omega_t $, $S \in \caB(\caH)$ as in \eqref{def:
K} and for $j=1,\ldots,n$
\begin{equation}
Z_{j}= \left\{ \begin{array}{ll} \eta^{1/2}_k(\om) V_{\omega,k} & \textrm{ if } t_j
\in \sigma_{\omega,k}
\\
- \eta^{1/2}_k(\om) V^{*}_{\omega,k} & \textrm{ if } t_j \in \tau_{\omega,k}
\end{array} \right.
\end{equation}
Finally, let \beq \caF:=L^2(\Om,\caH,\d \si) \simeq \caH \otimes
L^2(\Om,\d \si) \eeq Remark that in a natural way, we have $\caF
\simeq \caH \otimes \caR$ with $\caR$ as defined in
\eqref{fockspace}. Take $f \in L^2(\Omega,\mathcal{H},\d \si)$ and
define
\begin{equation}\label{solu}
(\mathbf{U}_t  f)(\xi)=\sum_{\sigma\subset \xi } \int_{ \Omega}
u_t(\sigma ,\tau)f((\xi
 \setminus \sigma)
\cup \tau )d\tau
\end{equation}
In \cite{maassen1}, one proves that this $\mathbf{U}_t$ is unitary
and that it solves the QSDE \eqref{dilation1}.

The unitary family $\mathbf{U}_t$, thus defined, is not a group,
but a so-called cocycle; physically this corresponds to an
interaction picture and it can be made into a group by multiplying
it with a well chosen `free evolution'.
%
%
Note that by taking each $V_{\omega,k}=0$ or $V_k=0$ in
(\ref{newv}) the subsystem decouples from the reservoir and
(\ref{solu}) reduces to
\begin{equation}\label{decoupled}
\mathbf{U}_t= 1 \otimes \mathbf{1}
\end{equation}
This follows since the kernel $u_t(\si,\tau)$ in \eqref{kernel}
vanishes except for $\si=\tau=\emptyset$ and $S$ reduces to $0$.

 Remark that in (\ref{dilation1}) or \eqref{solu}, the reservoirs are now not
only labeled by $k \in K$, as in the original physical picture,
but also by $\omega \in F$; each transition has its own
mathematical reservoir. To formulate our results, we also need to
specify the state. Define the one-dimensional vacuum projection
$\mathbf{1}_{\emptyset} \in \caB(\caF)$ \beq \label{def: vacuum}
(\mathbf{1}_{\emptyset}f)(\si) = \left \{
\begin{array}{ll}  f(\emptyset) &
\textrm{ when } \si =\emptyset \\
0 & \textrm{ when } \si \neq \emptyset  \end{array} \right.  \eeq
Our reference state is
\begin{equation}
 \rho \otimes \mathbf{1}_{\emptyset}   \textrm{  on  }
\mathcal{H}\otimes \mathcal{F}
\end{equation}
where $\rho$ is the unique stationary state of $\e^{t \caL}$, see
Remark \ref{rem: ergodicity}.

 Note that the state $\rho
\otimes \mathbf{1}_{\emptyset}$ is not invariant under the
dynamics, only its restriction to $\caH$ is invariant (see also
\eqref{dilation semigroup}). Hence, technically, it is quite
different from a NESS as in \eqref{def: ness}.

We will abbreviate the Heisenberg dynamics as
\beq j_t(\mathbf{G}) := \mathbf{U}^*_{t} \mathbf{G}
\mathbf{U}_{t},  \qquad \mathbf{G} \in \caB(\caH \otimes \caF)
\eeq with $\mathbf{U}_t$ as in (\ref{solu}).
 Let for each $k \in K, t \geq 0$, $\mathbf{N}^{t}_{k}  \in \caB(\caF)$ be
the energy operators  \beq \label{energy operators}
(\mathbf{N}^t_{k} f)(\si)= n^t_{k}(\si)f(\si) \eeq with $n^t_{k}$
as defined in \eqref{energy functions}.

 We also define a quantity which we interpret as the total energy
of subsystem plus reservoirs
\begin{equation}\label{energy}
\mathbf{N}^t := H_{S}+\sum_{\omega,k} \mathbf{N}^t_{k} , \qquad
\mathbf{N}^t \in \caB(\caH \otimes \caF)
\end{equation}
This interpretation is backed by Proposition \ref{conserv}.

These `energies' should be understood as renormalized quantities,
of which the (infinite) equilibrium energy of the reservoirs was
subtracted. This interpretation is confirmed by remarking that at
time $s=0$, these 'energies' equal $0$: for all continuous
functions $g$, \beq \Tr \left[ \mathbf{1}_{\emptyset}
g(j_{s=0}\mathbf{N}^t_{k}) \right] = \Tr \left[
\mathbf{1}_{\emptyset} g(\mathbf{N}^t_{k}) \right]=g(0) \textrm{
for all } k \in K, t \geq 0 \eeq

\subsubsection{Connection of the QSDE with the counting process}

The connection of the QSDE with the 'quantum trajectories' is
provided by the following lemma, which we will not prove. It can
be found for example in \cite{boutenkummerer,boutenguta} and it is
easy to derive starting from \eqref{solu} and remarking that \beq
\caW_t(\si) (\cdot) = u^*_t(\si,\emptyset)  \cdot
u_t(\si,\emptyset) \eeq

\begin{lemma}\label{lem: projections}
Let $E \subset \Omega $ be measurable (as for \ref{def: integration}). Denote by $\mathbf{1}_E$ the ortogonal
projection \beq \mathbf{1}_E : L^2(\Omega) \ra L^2(E) \eeq  and
recall $\mathbf{1}_{\emptyset}$ from \eqref{def: vacuum}. Then,
for all $A \in \caB(\caH)$
\begin{equation}\label{connection1}
\Tr_{\caF} \left[ \mathbf{1}_{\emptyset}  j_t(A \otimes
\mathbf{1}_E)  \right]=\int_{E} \d\sigma \, \caW_t(\sigma)A
\end{equation}
where $\Tr_{\caF}$ denotes the partial trace over $\caF $.
\end{lemma}\noindent

The formula \eqref{solu} actually defines a dilation of the
semigroup $\e^{ t \caL}$. To see this,
 take
$E =  \Omega$,  then \eqref{connection1} reads

\begin{equation}\label{dilation semigroup}
\Tr_{\caF} \left[ \mathbf{1}_{\emptyset}  j_t(A \otimes
\mathbf{1})  \right] =\int_{ \Omega} \d\sigma \, \caW_t(\sigma)A =
\e^{t \caL} A
\end{equation}
Another useful consequence of lemma \ref{lem: projections} is the
connection between the energy operators in \eqref{energy
operators} and the functions \eqref{energy functions}.
\begin{proposition} \label{prop: transcription}
Let $k_1,\ldots,k_{\ell} $, $t_1,\ldots,t_{\ell} $ and
$g_1,\ldots,g_\ell$ be finite ($\ell <\infty$) sequences of,
respectively, elements of $K$,  $\bbR^+$  and continuous
functions, and let
 $\mu \in \caT(\caH)$, then \beq \Tr \left[ (\mu \otimes
\mathbf{1}_{\emptyset}) \prod_{i=1}^{\ell} g_i (j_{t_i}(
\mathbf{N}^{t_i}_{k_i}  )) \right] = \mathbb{E}_{\mu}
\left[g_i(\prod_{i=1}^{\ell} n^{t_i}_{k_i}) \right]\eeq
\end{proposition}

\noindent Again, we do not give a complete proof and we refer to
\cite{boutenguta,boutenkummerer}. Proposition \ref{prop:
transcription} follows from lemma \ref{lem: projections} by using
that for all $t \geq s$ and $k \in K$ \beq j_t
(\mathbf{N}^s_k)=j_s (\mathbf{N}^s_k) \eeq and that the family $\{
\mathbf{N}^t_k \, \big|\, t
>0,k \in K \}$
is commutative.

\subsection{Results within the quantum picture}
First, we show that the energy (see \ref{energy}) is conserved.
\begin{proposition}\label{conserv}
Let $\mathbf{N}^t $  be as in (\ref{energy}). For all continuous
functions $g$:
\begin{equation}
\Tr \left[ (\rho \otimes \mathbf{1}_{\emptyset})
g(j_t(\mathbf{N}^t)) \right] = \Tr \left[ (\rho \otimes
\mathbf{1}_{\emptyset}) g(\mathbf{N}^t) \right]= \Tr \left[ \rho
 g(H_\sys) \right]
\end{equation}
\end{proposition}
The change of entropy in the environment up to time $t$ is
\begin{equation}\label{entpro1}
\mathbf{W}^t := \sum_{k \in K} \beta_k j_t(\mathbf{N}^t_{k})
\end{equation}
and its `steady state expectation' is the entropy production. Our
main result is a fluctuation theorem for $\mathbf{W}^t$.

\begin{proposition}\label{ft}
Assume {\bf A1}. Let $\mathbf{W}^t$ be defined as in
(\ref{entpro1}). There is an open set $\caU \in \bbC$ containing
the real line, $\bbR \subset \caU$, such that for all $\ka \in
\caU$, the limit
\begin{equation}\label{rate function}
\hat{e}(\kappa) := \lim_{t \uparrow +\infty} \frac{1}{t}\log \Tr
\left[(\rho \otimes \mathbf{1}_{\emptyset}) e^{-\kappa
\mathbf{W}^t} \right]
\end{equation}
exists and the function $\ka \to \hat{e}(\kappa)$ is analytic on
$\caU$. Let $e(\kappa)$ by defined as in (\ref{rate function
clas}). Then,
\begin{equation} \label{symmetryhat}
\hat{e}(\kappa)=e(\kappa)
\end{equation}
on $\caU$ and thus all statements in Proposition \ref{thm: ft}
carry over to $\hat{e}(\kappa)$.
\end{proposition} \noindent
From  $\rho \otimes \mathbf{1}_{\emptyset}$ we deduce probability
measures $\mathbb{T}_t$ on $\mathbb{R}$. Let $A \subset
\mathbb{R}$ be measurable, then
\begin{equation}
\mathbb{T}_t(A)=\Tr \left[ (\rho \otimes \mathbf{1}_{\emptyset})
\mathbf{1}_{A} (\mathbf{W}^t) \right]
\end{equation}
where $\mathbf{1}_{A} (\mathbf{W}^t)$ is the spectral projection
on $A$ associated to $\mathbf{W}^t$. Via Legendre-transformation
(\ref{symmetry}) implies
\begin{equation}\label{entproform}
-\lim_{t \uparrow +\infty}\frac{1}{t}
\log{\frac{d\mathbb{T}_t(-a)}{d\mathbb{T}_t(a)}}=a
\end{equation}
which is (\ref{entproform2}).

 In the same way as in Proposition
\ref{ft}, Propositions \eqref{thm: green kubo} and \eqref{thm:
strict positivity} carry over the quantum picture; for
concreteness we give the analogue of Proposition \eqref{thm: green
kubo}.

\begin{proposition}\label{prop: qt greenkubo}
Assume {\bf A1} and fix some $\be > 0$. Let for $k,k' \in K$:
\begin{equation}
\tilde{L}_{k,k'}(\be) := \frac{\partial}{\partial
\beta_{k'}}\lim_{t \uparrow +\infty} \frac{1}{t} \Tr \left[ (\rho
\otimes \mathbf{1}_{\emptyset})\mathbf{N}^t_k )\right]
\Big|_{\be_1=\ldots=\be_{|K|}=\be}
\end{equation}
and similarly the time-reversed coefficient
$\tilde{L}^{\theta}_{k,k'}$, obtained by starting with
$H_\sys^{\theta}$ and $V_k^{\theta}$.
\begin{equation} \label{eq: quantum modified green kubo}
\tilde{L}_{k,k'}(\be)+ \tilde{L}^{\theta}_{k,k'}(\be)=  \beta
\lim_{t \uparrow +\infty}  \frac{1}{t} \Tr \left[(\rho \otimes
\mathbf{1}_{\emptyset}) \mathbf{N}^t_{k} \mathbf{N}^t_{k'} \right]
\end{equation}
 If also {\bf A3} holds, then
\begin{equation} \label{eq: quantum green kubo}
\tilde{L}_{k,k'} = \frac{1}{2}\beta \lim_{t \uparrow +\infty}
\frac{1}{t} \Tr \left[(\rho \otimes \mathbf{1}_{\emptyset})
\mathbf{N}^t_{k} \mathbf{N}^t_{k'} \right]
\end{equation}
with Onsager reciprocity
\begin{equation} \label{eq: quantum onsager reciprocity}
\tilde{L}_{k,k'} = \tilde{L}_{k',k}
\end{equation}

\noindent Recall $L_{k,k'}$ from Proposition \ref{thm: green kubo}. Then for
all $k \in K$, \beq \tilde{L}_{k,k'}=L_{k,k'}\eeq


\end{proposition}

Remark that Propositions \ref{ft} and \ref{prop: qt greenkubo}
follow immediately from Propositions   \ref{thm: ft} and \ref{thm:
green kubo} by application of Proposition \ref{prop:
transcription}.

\section{Discussion} \label{sec: discussion}

\subsection{Entropy production for Markov
processes}\label{clasweak}
 It is well known that the weak coupling generator
is `classical' in the sense that the commutant algebra $\caA_{cl}
:= \big\{A \in \caB(\caH)  \, \big| \,  [A,H_\sys]=0 \big\}$ is
invariant. In case the Hamiltonian $H_\sys$ is non-degenerate and
only then, $\caA_{cl}$ is a commutative algebra. Then we can
construct a Markov process with state space $\Lambda$ which is the
restriction of (the dual of) the semigroup $e^{t\mathcal{L}}$ to
$\caA_{cl} \simeq \caC(\Lambda)$. Loosely speaking, let $\rho$ be
the stationary state, $\Omega^t := \Lambda^{[0,t]} $ the pathspace
up to time $t$, and $\caP^t_{\rho}$ the pathspace measure
(starting from $\rho$) of this Markov process. The time reversal
operation $\Theta$ acts on
$\Omega^t$ as  $(\Omega \xi)(u)=\xi(t-u)$ for $\xi \in \Om^t $ and $0 \leq u \leq t$.\\

For such Markov processes describing a nonequilibrium dynamics, we
dispose of a general strategy for identifying the entropy
production. It turns out in a lot of interesting cases
\cite{maesirreversible,maesnetocnyverscheure,maes99} that
\begin{equation}\label{clasft}
\log{\frac{d\caP^t_{\rho}}{d\Theta
\caP^t_{\rho}}}(\xi)=S_t(\xi)+\caO(1)
\end{equation}
where $S_t(\xi)$ is the random variable that one physically
identifies as the entropy production. The second term in the
righthand side is non-extensive in time. The algorithm allows to
derive (\ref{entproform2}) from (\ref{clasft}).\\
Since we also have a Markov generator, we can apply the same
scheme to our setup.\footnote{Very recently, a paper
\cite{espositomukamel} appeared where exactly this is done: one
derives a fluctuation theorem for
$\log{\frac{d\caP^t_{\rho}}{d\Theta \caP^t_{\rho}}}$ as in
\eqref{clasft}. Since the authors consider mainly examples
involving one reservoir, they do not run into the difficulty
described here.}  To evaluate the result, we however need a
physical notion of entropy production in our model. As mentioned
earlier, such a notion is rather unambiguous here, see also
\cite{lebowitzspohn1,jaksicpillet3}:
\begin{equation}\label{spohns mean}
\langle\textrm{current into $k$'th reservoir} \rangle = \Tr \left[
 \rho \caL_k H_\sys \right]
\end{equation}
But the mean entropy production based on these currents is {\bf
not} equal to the expectation value of (\ref{clasft}):
\begin{equation}\label{inequality}
\sum_{k \in K} \beta_k \Tr \left[
 \rho \caL_k H_\sys \right]  \neq  \lim_{t
\uparrow \infty} \frac{1}{t} \mathbb{E}_{\caP^t_{\rho}} \left[
\ln{\frac{d\caP^t_{\rho}}{d\Theta \caP^t_{\rho}}} \right]
\end{equation}
For example, take two reservoirs $(k=L (\textrm{left}) ,R
(\textrm{right}))$ and let $\refl : \caB(\caH) \to \caB(\caH)$
stand for the involution which models left-right reflection.
Assume that $H_\sys$ is non-degenerate and that for all $x \in
\bbR$, \beq \refl H_\sys = H_\sys ,\qquad    \refl V_{L} =  V_{R}
,\qquad \e^{\be_L x/2}\eta_L(x)=\e^{\be_R x/2}\eta_R(x) \eeq Hence
all parameters are left-right symmetric, except the inverse
temperatures $\be_L,\be_R$. (Actually, these assumptions are
inconsistent; if $H_\sys$ is left-right symmetric, then it must be
degenerate. However, one can introduce an arbitrarily small
symmetry breaking which will generically lift the degeneracy, such
that the our reasoning still applies.) One  checks that
\begin{eqnarray} \label{eq: nonextensive radon}
\log{\frac{d\caP^t_{\rho}}{d\Theta
\caP^t_{\rho}}}(\xi)&=&\log{\frac{\mathop{\sum}\limits_{e_0 \in
A_0 \subset \sp H_\sys }\rho \left[1_{e_0}(H_\sys) \right]
}{\mathop{\sum}\limits_{e_m \in A_m \subset \sp H_\sys }\rho
\left[1_{e_m}(H_\sys) \right]}}+ \sum_{i=1}^{m} \log{
\frac{\e^{-\frac{\be_L \omega_i}{2}}+\e^{-\frac{\be_R
\omega_i}{2}}}{\e^{\frac{\be_L \omega_i}{2}}+\e^{\frac{\be_R
\omega_i}{2}}}} \nonumber\\& =& \caO(1)-
(\be_L+\be_R)\sum_{i=1}^{m}\om_i \end{eqnarray} where the sets
$A_0,A_m$ and the sequence $\om_{i=1}^m$ of energy jumps are
derived from $\xi$, and moreover $|\sum_{i=1}^{m} \om_i| \leq \|
H_\sys \|$.  This means that in this particular left-right
symmetric case, \eqref{eq: nonextensive radon} is bounded,
independently of $t$ for every $\xi$, and hence the RHS of
\eqref{inequality} vanishes, which disqualifies it as ``entropy
production".

 This trivial remark shows that it is not enough to look
at the semigroup $\e^{t\caL}$ to identify the entropy production.
Instead we use more input; we certainly use the fact that $\caL
=\sum_{k \in K} \caL_k$ where the index $k$ runs over the
different reservoirs but moreover, with the unraveling of the
generator, Section \ref{unravel}, comes an {\bf intuitive}
interpretation of the various terms. That can be contrasted with
results by V.~Jak${\check {\mathrm s}}$i\'{c} and C.~A.~Pillet,
where one actually {\bf proves} that quantities like $\Tr \left[
 \rho \caL_k H_\sys \right]$, cfr.\ \eqref{spohns mean} are limits of
currents in the original microscopic Hamiltonian model. Of course,
we take care that our choices are consistent with that result.
However, for the higher-order fluctuations, we do not know; we
just make a choice which looks very natural. At present, we do not
give arguments that for a class of reasonable functions $g$
\begin{equation}\label{fluctuation}
\bbE_{\rho} \left[ g (w^t) \right]
\end{equation}
is indeed the limit of some fluctuation of dissipated heat in the
microscopic model. (Although \cite{derezinskideroeck2} points in that direction, see also point 2 in section \ref{sec: connection})\\
Another choice for the higher order fluctuations is discussed in
Section \ref{sec: integrated currents}. It is exactly here that
lies the role of the dilation with quantum stochastic evolutions.
If one takes that quantum model as a starting point, then one can
{\bf derive} that (\ref{fluctuation}) is a fluctuation of the
dissipated heat. To our knowledge, that is the only quantum model
in which one can study the fluctuations of the dissipated heat.

On the other hand, one can also make a classical dilation of the
semigroup and in fact, this is exactly what we do in Section
\ref{unravel}. Yet, there is a technical difference between the
cases of degenerate and non-degenerate system Hamiltonians
$H_\sys$. If $1_e(H_\sys)$ is one-dimensional for $e \in \sp
H_\sys$, and in addition, for a nonzero $\om \in F$, $e$ is the
unique element of $\sp H_\sys$ such that $e-\om \in \sp H_\sys$,
then we have the following form of Markovianness: If a $\si \in
\Om$ contains $\om$, i.e.\ \beq \si=\si_0 \tau \si_1, \qquad \si_0
\in \Om_{t_0},\si_1 \in \Om_{t_1}, \tau=(t_0,\om,k) \textrm{ for
some } k \in K,\,t_0,t_1 \geq 0 \eeq then
\begin{eqnarray} \d \bbP_{\rho} (\si) &=& \Tr \left[ \rho  \caW_{t_0+t_1}(\si) 1 \right] \nonumber \\
& =& \Tr \left[ \rho
\caW_{t_0}(\si_0 \tau) 1 \right] \d \si_0 \d \tau \times \Tr \left[ 1_e (H_\sys) \caW_{t_1}(\si_1)1 \right] \d \si_1  \nonumber \\
&=& \d \bbP_{\rho} (\si_0 \tau) \, \d \bbP_{1_e (H_\sys)} (\si_1)
\end{eqnarray}  In words, a
one-dimensional spectral subspace erases memory. That does not work in the degenerate case.

\subsection{Integrated currents within the semigroup approach }
\label{sec: integrated currents} Starting from (\ref{spohns mean})
one could define the integrated currents $\hat{J}_{k,t}
\in \caB(\caH)$ as
\begin{equation}\label{Spohns current}
\hat{J}_{k,t}= \int_0^{t}\d u \, \e^{u \caL}(\caL_k(H_\sys))
\end{equation}
and study their fluctuations. One can ask whether these
fluctuations coincide with these in our model? The answer is
partially positive because
\begin{proposition}\label{current-current}
Take for all $k \in K: \, \beta_{k}=\be$ for a certain $\be$ and
let $\rho_{\be}$ be the stationary state for $\e^{t \caL}$ as in
Remark \ref{rem: canonical}. For all $k,k' \in K$ and all $u \geq
0$,
\begin{equation}\label{currenteq}
\frac{\partial^2}{\partial v_1 \partial v_2} \mathbb{E}_{\rho_{\beta}} [
n^{v_1}_{k'}n^{v_2}_{k} ]\big|_{v_1=0,v_2=u}=- \Tr \left[
\rho_{\beta} \caL_{k'}(H_\sys) \e^{u \caL}(\caL_{k}(H_\sys))
\right]
\end{equation}
\end{proposition}\noindent
which gives a relation between Proposition \ref{thm: green kubo}
and the Green-Kubo relation in \cite{lebowitzspohn2}. Also the
averages coincide, leading to \eqref{spohns mean}. However it is
{\bf not} true that for a reasonable class of functions $g$
\begin{equation}
 \mathbb{E}_{\rho}
[g(n^t_{k}) ]=\Tr \left[ \rho g(\hat{J}_{k,t}) \right]
\end{equation}
So the mean entropy production and the Green-Kubo-formula can
correctly be expressed in terms of the operators $\hat{J}_{k,t}$,
but higher order fluctuations of the dissipated heat cannot.
\subsection{Connection to microscopic dynamics}\label{sec: connection}
We know of three derivations in the literature of
the stochastic evolution \eqref{solu} or \eqref{dilation1} from a microscopic setup:\\

\noindent {\bf 1 Stochastic Limit}\\
Accardi \textit{et al}. prove in \cite{accardifrigerio} that the
weak coupling limit can be extended to the total evolution of
subsystem observables. Let $\mathcal{U}^{\lambda}_{t}$ be the
evolution (in the interaction picture) on the total system with
$\lambda$ the coupling between subsystem and reservoirs. Then, in
a certain sense,
\begin{equation}
\caU^{\lambda}_{-t/\lambda^2} (S \otimes
\mathbf{1})\caU^{\lambda}_{t/\lambda^2}
\mathop{\rightarrow}\limits_{\lambda \downarrow 0}
\mathbf{U}^*_{t} (S \otimes \mathbf{1})\mathbf{U}_{t}
\end{equation}
whereas the traditional weak coupling limit only speaks about
convergence in expectation of the left-hand side. The unitary
$\mathbf{U}_t$ is the solution of (\ref{dilation1}).\\

\noindent{\bf 2 Stochastic Limit Revisited}\\
In \cite{derezinskideroeck2}, the approach of
\cite{accardifrigerio} (mentioned above) was simplified. By
introducing a unitary map $J_{\la}$ acting on the reservoirs, we
get for all continuous functions $g$
\begin{equation}
\mathrm{s}-\lim_{\la \downarrow 0} J^*_{\la}
\caU^{\lambda}_{\lambda^{-2}t} J_{\la} = \mathbf{U}_{t} , \qquad
\mathrm{s}-\lim_{\la \downarrow 0} J^*_{\la} g(H_{\res_k})
J_{\la}= g(\mathbf{N}_k)
\end{equation}
where $\mathrm{s}-\lim$ denotes strong operator convergence and
$H_k$ is the generator of the dynamics in the uncoupled $k$'th
reservoir. This suggests that one can study the fluctuations of
the reservoir energies by looking at the number operators
$\mathbf{N}_k$ in the model reservoirs, exactly as we do in the
present paper.\\

\noindent{\bf 3 Repeated Interactions}\\
In \cite{attalpautrat}, Attal and Pautrat describe a subsystem
with Hilbertspace $\mathcal{H}$ interacting repeatedly for a time
$h$ with a small reservoir with Hilbertspace $\mathcal{R}$. After
each time $h$, $\mathcal{R}$ is replaced by an identical copy.
This procedure ensures that at any time, the subsystem sees a
`fresh' reservoir. In the limit $h \rightarrow 0$ the dynamics (in
the interaction picture) converges in a certain sense to the
solution of a QSDE. One can choose a particular QSDE by tuning the
parameters of the interaction. Assume that
\begin{equation}
\mathcal{R}= \otimes_{\omega,k} \mathcal{R}_{\omega,k}
\end{equation}
Each $\mathcal{R}_{\omega,k}$ is 2-dimensional with basevectors
$(\theta,\omega)$. Define $a_{\omega,k}$ on
$\mathcal{R}_{\omega,k}$ by
\begin{equation}
a_{\omega,k}(\omega)= \theta ,  \qquad a_{\omega,k}\theta= 0
\end{equation}
Choose the dynamics on $\mathcal{H} \otimes \mathcal{R}$ as
$e^{-itH(h)}$ for $0\leq t \leq h$ with
\begin{equation}
H(h)=H_{f}+ \frac{1}{\sqrt{h}}\sum_{\omega,k}
\big(V_{\omega,k}a^*_{\omega,k}+ V^*_{\omega,k}a_{\omega,k} \big)
\end{equation}
Then, through the limiting procedure of \cite{attalpautrat},
equation (\ref{dilation1}) obtains.
%
%
%

%
%
%
%

%
%
%
%
%
%

\section{Proofs}
\subsection{Proof of Lemma \ref{prob} }
From
\begin{equation} \label{normalization measure}
 \int_{\Omega_t}d\sigma \,
\mathcal{W}_t(\sigma)1  =e^{t\caL}1=1,  \qquad \forall \si \in
\Om: \, \mathcal{W}_t(\sigma)1 \geq 0
\end{equation}
for all $t \geq 0$, it follows that $(\mathbb{P}_{\mu,t})_{t \in \bbR^+}$ is indeed a family
of probability measures for all $\mu \in \caT(\caH)$. Further, for $s,u \geq 0$, we have
\beq
\caW_t(\si)\caW_t(\tau)=\caW_t(\si \tau),  \qquad \si \in \Om_s, \tau \in \Om_u
\eeq
Together with \eqref{normalization measure}, this yields consistency of the family $(\mathbb{P}_{\mu,t})_{t
\in \bbR^+}$.

%
%
%

%
%
%
%
\subsection{Proof of Proposition \ref{thm: ft}}
Define for $\overline{\kappa} \in \bbC^{|K|}$ and $t >0$, \beq
M_{t,\overline{\ka}}: \Omega \mapsto \bbC, \quad
M_{t,\overline{\ka}}(\si) = \sum_{k \in K} \ka_k \be_k
n^t_{k}(\si) \eeq Our results  rely on the following lemma
\begin{lemma}\label{lem: analytic}
Assume {\bf A1} and let $\mu \in \caT(\caH)$. There is an open set
$\caU \subset \bbC^{|K|}$, with $\bbR^{|K|} \subset \caU$  such that
 \beq
\e(\overline{\ka}):=\lim_{t \uparrow +\infty}
\frac{1}{t}\log{\mathbb{E}_{\mu} \left[ \e^{M_{t,\overline{\ka}}}
\right]} \eeq is an analytic function on $\caU$ which does not
depend on $\mu$. Moreover, for any sequence $k_1,\ldots, k_{\ell}
\in K$, \beq \label{eq: derivatives converge} \lim_{t \uparrow
+\infty} \frac{\partial}{
\partial \ka_1} \ldots \frac{\partial}{ \partial \ka_{\ell}}
\frac{1}{t}\log{\mathbb{E}_{\mu} \left[ \e^{M_{t,\overline{\ka}}}
\right]}= \frac{\partial}{ \partial \ka_1} \ldots \frac{\partial}{
\partial \ka_{\ell}}e(\overline{\kappa}) \eeq uniformly on
compacts.
\end{lemma}
\begin{proof}
We apply the generalized Perron-Frobenius Theorem \ref{app:
perron} of the Appendix with
 \beq \label{the map La} \La =
\int_{\Omega}\d \si \, \caW_{t_{\ep}} (\si)
\e^{M_{t_{\ep},\overline{r}} (\si)} \eeq for well chosen $t_{\ep}$
and $\overline{r} \in \bbR^{|K|}$.
Since for $\overline{r} \in \bbR^{|K|}$, $ M_{t_{\ep},\overline{r}}$ is a real function, the map $\La$ is completely positive as a linear combination of completely positive maps with positive coefficients.\\
Below we choose $t_{\ep}$ so as to satisfy the
non-degeneracy requirement \ref{cond: positivity} of the Appendix.\\
By faithfulness of the stationary state $\rho$, \beq \ep :
=\inf_{0<P \in \caB(\caH),P^2=P } \Tr \left[\rho P \right]>0 \eeq
Since the semigroup is ergodic, it follows that there is $t_{\ep}$
such that for all $t > t_{\ep}$,
\beq \sup_{\mu \in \caT(\caH)} \| \rho  - \e^{t \caL^* } \mu \|
\leq \frac{\ep}{3(\dim \caH)^2} \eeq
Since $\| \caL \|, \| \caL_0 \|  < + \infty $, with $\| \cdot \|$
being the operator norm in $\caB(\caB(\caH))$, the Dyson expansion
\eqref{def: dyson} is absolutely convergent. Hence, we can find $n
\in \bbN$ such that
 \beq    \|   \int_{|\si| \leq n } \d \si \, \caW_{t_{\ep}} (\si)  -
\e^{t_{\ep} \caL} \| \leq \frac{\ep}{3 \dim \caH} \eeq
Let $m:= \inf_{|\si| \leq n} M_{t_{\ep},\overline{r}} (\si)$. For
each $\overline{r} \in \bbR^{|K|}$, decompose
\begin{eqnarray} \label{deformedgen}
&& \int_\Om \d \si \, \caW_{t_\ep}(\si) \e^{M_{t_{\ep},\overline{r}}}= \e^{
m} \int_{|\si| \leq n } \d \si \, \caW_{t_{\ep}} (\si) \\&& +
\int_{|\si| \leq n } \d \si \, \caW_{t_{\ep}} (\si)
(\e^{M_{t_{\ep},\overline{r}} (\si)}-\e^{m} )  + \int_{|\si| > n }
\d \si \, \caW_{t_{\ep}} (\si) \e^{M_{t_{\ep},\overline{r}} (\si)}
\nonumber
\end{eqnarray}
and for each pair of non-zero projections $P \neq 0,P' \neq 0 \in
\caB(\caH)$, we have
\begin{eqnarray} \label{deformedgen 2}
&& \Tr \left[ P \int_{\Omega} \caW_{t_\epsilon}(\si)
\e^{M_{t_\epsilon,\overline{r}}}\, P' \right] \geq \e^{m}
\int_{|\si| \leq n } \d \si  \Tr \left[ P \caW_{t_{\ep}} (\si) P' \right]\nonumber \\
&&\geq   \e^{m} \big( \Tr \left[ P \e^{t_{\ep} \caL} P'  \right] - \frac{\ep}{3}
\big)  \nonumber \\
&& \geq \e^{m} \big( \Tr \left[ P' \rho\right] - \frac{\ep}{3} -
\frac{\ep}{3} \big)   \geq  \e^{m} \frac{\ep}{3}
\end{eqnarray}
This shows that one can apply Theorem \ref{app: perron} with $\La$
as in \eqref{the map La}. Call the dominant eigenvalue of $\La$,
$\la(\overline{r},t_{\ep})$ and the corresponding strictly
positive eigenvector
 $v (\overline{r})$. Remark that for each
$\overline{\ka} \in \bbC^{|K|}$ and $t \in \bbR^+$,  \beq
\label{using gen kappa} \int_\Om \d \si \, \caW_t(\si)
\e^{M_{t,\overline{\ka}}(\si)} = \e^{t\caL_{\overline{\ka}}}  \eeq
where
\begin{equation} \label{gen kappa}
\mathcal{L}_{\overline{\ka}}(\cdot)=\caL_0(\cdot)+\sum_{\omega,k}
\eta_k(\omega) \e^{-\ka_k \be_k \om} V_{\omega,k}^{*}\cdot V_{\omega,k}
\end{equation}
This follows by comparing the Dyson expansions (in the same sense
as for \eqref{def: dyson}) corresponding to the left-hand and the
right-hand side of \eqref{using gen kappa}. As a consequence, for
all $\overline{r} \in \bbR^{|K|}$, $\mathcal{L}_{\overline{r}}$
has a non-degenerate maximal eigenvalue $\la(\overline{r})
=\frac{1}{t_{\ep}} \ln \la(\overline{r},t_{\ep})$ corresponding to
the eigenvector $v (\overline{r}) $. Since $v (\overline{r}) $ is
strictly positive, we have $\Tr \left[ v (\overline{r})
\right]>0$, and, for any $\mu \in \caT(\caH)$, $\Tr \left[ v
(\overline{r}) \mu \right]
>0$. This implies
\begin{equation}\label{rate f equals ev}
\lim_{t \uparrow +\infty}  \frac{1}{t}\log \Tr \left[ \mu
\e^{t\caL_{\overline{r}}} 1 \right]    = \la(\overline{r})
\end{equation}
and hence, again by \eqref{using gen kappa}
\beq
e(\overline{r})=\la(\overline{r})
\eeq

Since for all $\overline{\ka} \in \bbC^{|K|}$,
$\caL_{\overline{\ka}}$ depends analytically on $\overline{\ka}$,
perturbation theory for isolated eigenvalues gives us for all
$\overline{r} \in \bbR^{|K|}$ an open set $U_{\overline{r}} \ni
\overline{r}$ such that for all $ \overline{\ka} \in
U_{\overline{r}}$ :
\begin{enumerate}
\item{There is a unique $\la(\overline{\ka}) \in \sp
\caL_{\overline{\ka}}$ such that \beq   \inf \{ \Re
\la(\overline{\ka}) -|p| \, \big|\, p \in \sp
\caL_{\overline{\ka}} \setminus \la(\overline{\ka}) \} >0   \eeq }
\item{ The eigenvector $v(\overline{\ka})$, corresponding to $\la(\overline{\ka})$ satisfies
\beq \label{cond: nonzero} \inf_{\mu \in \caT(\caH)} \left(\Re \Tr
\left[ \mu v(\overline{\ka})\right] \right)
>0    \eeq }
\end{enumerate}
It follows that \eqref{rate f equals ev} holds for all
$\overline{\ka} \in \mathop{\cup}\limits_{\overline{r} \in
\bbR^{|K|}} U_{\overline{r}}$,
\begin{equation}\label{complex rate f equals ev}
e(\overline{\ka})= \lim_{t \ra \infty}  \frac{1}{t}\log  \Tr
\left[ \mu \e^{t\caL_{\overline{\ka}}} 1 \right] =
\la(\overline{\ka})
\end{equation}
Summarizing, we have  for all $\overline{r} \in \bbR^{|K|}$ and
$\mu \in \caT(\caH)$ a family of analytic functions \beq
 F(t,\overline{\ka}):= \frac{1}{t}\log{ \Tr \left[ \mu  \e^{t \caL_{\overline{\ka}}}1 \right] }\eeq
 converging pointwise in $U_{\overline{r}}$ to the function $e(\overline{\ka})$ as $t \uparrow +\infty$.

We recall Montel's Theorem, see e.g. p. 153 of \cite{conway}:

\begin{theorem}\label{thm: montel}
Let $G \subset \bbC$ be open and let $(f_n)_{n \in \bbN}$ be a
sequence of analytic functions $G \mapsto \bbC$, then $(f_n)_{n
\in \bbN}$ contains a uniformly convergent on compacts subsequence
iff the set $(f_n)_{n \in \bbN}$  is locally bounded, i.e., that
for each $z \in G $ there is a $r >0$ and $M >0$, such that \beq
   |z'-z | \leq r   \Rightarrow \forall n \in \bbN: \, |f_n(z')| \leq M
\eeq
\end{theorem}

%
%

For all $\overline{r} \in \bbR^{|K|}$, the family
$(F(t,\overline{\ka}))_{t \geq t_0}$ is locally bounded on
$U_{\overline{r}}$ for large enough $t_0 \geq0 $. This follows
from analyticity of $\caL_{\overline{\ka}}$ and from the condition
\eqref{cond: nonzero}. Consequently, one can apply Theorem
\ref{thm: montel} for each component of $\overline{\kappa}$
separately. A standard result, e.g.\ Theorem 2.1, p.\ 151 in
\cite{conway} states that the uniform limit of a sequence of
analytic functions is analytic and that all derivatives converge.
Since this generalizes to the multidimensional variable
$\overline{\kappa}$, e.g.\ by Hartog's theorem, Lemma \ref{lem:
analytic} is proven.
\end{proof}

%
%
%
%
Referring again to the representation \eqref{ordered}, we
introduce for $\si \in \Omega$ the factor \beq \eta (\si):=
\prod_{i=1}^{| \si|} \eta_{k_i}(\omega_i)\eeq Recall the
definition of $S$ in \eqref{expressing L0 in K}, introduce the
time-reversed maps $\caL^{\theta}_0$ and, for $t \geq 0$,
$\caW^{\theta}_t$ (i.e., these maps are derived from
$H^{\theta}_S$ and $V_k^{\theta}$) and remark, (see also
\eqref{expressing L0 in K})
\beq \label{the reversed free generator}  \e^{t\caL_0} (\cdot) =
\e^{t S} \cdot \e^{t S^*} \qquad \e^{t\caL_0^{\theta}} (\cdot) =
\e^{t TS^* T} \cdot \e^{t TS T} \eeq Define the operation $\theta_t$
on $\Om_t$ as \beq \label{def: time reversal}\theta_t
(\om_1,k_1,t_1;\ldots ;\om_n,k_n,t_n ) := (-\om_n,k_n,t-t_n ;
\ldots ; -\om_1,k_1,t-t_1 ) \eeq
Calculate
\begin{eqnarray} \label{eq: path reversal}
&& \eta^{-1} (\si) \Tr \left[ \caW_t(\si) 1  \right]  \\
&=&  \Tr \left[  \ldots   V^*_{\om_i,k_i}\e^{(t_{i+1}-t_i)S} V^*_{\om_{i+1},k_{i+1}} \ldots   V_{\om_{i+1},k_{i+1}}      \e^{(t_{i+1}-t_i)S^*}  V_{\om_i,k_i}  \ldots  \right]   \nonumber     \\
&=&  \Tr \left[  \ldots TV_{\om_{i+1},k_{i+1}}  TT    \e^{(t_{i+1}-t_i)S^*} TT V_{\om_i,k_i}T    \ldots            T  V^*_{\om_i,k_i}  TT\e^{(t_{i+1}-t_i)S}  TT  V^*_{\om_{i+1},k_{i+1}} T  \ldots   \right]  \nonumber\\
&=&   \Tr \left[   \ldots TV^*_{-\om_{i+1},k_{i+1}}  T    \e^{(t_{i+1}-t_i)TS^*T} T V^*_{-\om_i,k_i}T    \ldots            T  V_{-\om_i,k_i}  T\e^{(t_{i+1}-t_i)TST}  T  V_{-\om_{i+1},k_{i+1}} T  \ldots      \right]    \nonumber\\
&=& \eta^{-1} (\theta_t \si)        \Tr \left[
\caW_t^{\theta}(\theta_t \si) 1 \right]  =     \eta^{-1} ( \si)
\e^{w_t(\si)}    \Tr \left[ \caW_t^{\theta}(\theta_t \si) 1  \right]
\nonumber
\end{eqnarray}
In the last equality the KMS-condition \eqref{KMS condition} was
used. The previous equalities follow from cyclicity of the trace,
$TT=1$, $V^*_{\om,k}=V_{-\om,k} $ and \eqref{the reversed free
generator}. Using \eqref{eq: path reversal}, we calculate by
change of integration variables (putting $I := \frac{1}{\dim \caH}
\in \caT(\caH)$,
\beq \label{eq: change of integration variables}
 \bbE_{I} \left[ \e^{M_{t,\overline{\ka}}}  \right]= \int_\Om \d \si \, \Tr \left[ I\caW_t(\si) 1 \right] \e^{M_{t,\overline{\ka}} (\si)}
= \int_\Om \d  \si \, \Tr \left[ I\caW^{\theta}_t(\si) 1 \right]
 e^{-w^t(\si)} \e^{-M_{t,\overline{\ka}}
(\si)}
\eeq
Since in the limit $t \uparrow \infty$, one can replace the
initial state $I$ by $\rho$, as in \eqref{complex rate f equals
ev}, the formula \eqref{eq: change of integration variables}
yields for all $\overline{\ka} \in \caU$ as in Lemma \ref{lem:
analytic}. \beq \label{eq: general ft} e(\overline{\ka})
=e^{\theta}(1-\overline{\ka}) \textrm{   with } 1-\overline{\ka}:=
(1-\ka_1, \ldots, 1-\ka_{|K|}) \eeq
Finally, Proposition \ref{thm: ft} follows from \eqref{eq: general
ft} by putting for some $\ka \in \bbC$, \beq \ka_i := \ka   \quad
i=1,\ldots,|K| \eeq thus obtaining $M_{t,\overline{\ka}}=\ka w^t$

\subsection{Proof of Proposition \ref{thm: strict positivity}}

The nonnegativity of the entropy production follows from
Proposition \ref{thm: ft} by Jensen's inequality. To get the
strict positivity from {\bf A2}, we first need to introduce more
notation.\\

Let $\rho$ be the unique stationary state of $\e^{t \caL}$. We
decompose the states $\rho$ and $ T \rho T$ in one-dimensional
unnormalized states as \beq \label{eq: decomp state1}
\rho^{\star}=\sum_{i \in D}\rho^{\star}_i  ,\quad \rho^{\star}_i
\rho^{\star}_j =\de_{i,j}\|\rho^{\star}_i \| \rho^{\star}_i ,
\quad \rho^{\star}_i
>0 , \quad i,j \in D\eeq where
$\rho^{\star}$ can stand for $\rho$ or $ T \rho T$ and
$D:=\{1,\ldots,\dim \caH \}$. The decomposition \eqref{eq: decomp
state1} differs from the spectral decomposition when $\rho$ is
degenerate. Remark that there is an arbitrariness in labeling the
unnormalized states, as well as a possible arbitrariness stemming
from degeneracies in $\rho^{\star}$. We partially fix this
arbitrariness by asking that \beq T (T\rho^{\star}T)_j T=
\rho^{\star}_j \eeq This is always possible because the set $
T(T\rho^{\star}T)_j T , j \in D$ satisfies all the requirements of
\eqref{eq: decomp state1} as a decomposition of $\rho^{\star}$.
Let $\tilde{\Omega}_t= \Om_t \times D \times D$ for a $t \geq 0$
and define the measure $\tilde{\bbP}_t$ by (letting $g$ be a
measurable function) : \beq \label{def: tilde
measure}\int_{\tilde{\Om}_t} \d \tilde{\bbP}_t(\tilde{\si})
g(\tilde{\si})= \sum_{i,j} \int_{\Om_t} \d \si \Tr \left [\rho_i
\caW_t(\si)\frac{(T\rho T)_j}{\| (T\rho T)_j\|} \right ]
g(\si,i,j),   \qquad \tilde{\si} =(\si,i,j) \in \tilde{\Om}_t
 \eeq
where it is understood that $\si \in \Om_t$ and $i,j \in D$. In
the rest of this section we will use this notation without further
comments. Positivity of $\tilde{\bbP}_t$ is obvious and
normalization follows by
\begin{eqnarray} \label{eq: check normalization}&&\sum_{i,j}\int_{\Om_t} \d \si \Tr \left [
\rho_i  \caW_t (\si)\frac{(T\rho T)_j}{\| (T\rho T)_j\|} \right ] \nonumber\\
&&=\sum_{i}\int_{\Om_t} \d \si \Tr \left [ \rho_i \caW_t (\si)1 \right ] =\int \d
\bbP_{\rho,t} (\si)=1
\end{eqnarray}

We call $\tilde{\bbP}^{\theta}_t$ the measure, constructed as
above, but with $\caW_t^{\theta}$ replacing $\caW_t$. Remark that
this is {\bf not} the measure one would obtain by starting from
$H^{\theta}_S, V^{\theta}_k$ instead of $H_\sys, V_k$, because
then one would also replace $\rho$ in \eqref{def: tilde measure}
by
$\rho^{\theta}$, the stationary state of $\caL^{\theta}$.\\
Define again the operation $\theta_t$ on $\tilde{\Omega}_t$ as
\beq \theta_t (\si,i,j) = (\theta_t \si,j,i) \eeq where the action
of $\theta_t$ on
$\Om_t$ was defined in \eqref{def: time reversal}.\\
Consider the function \beq \label{def: S}S^t: \tilde{\Omega}_t
\mapsto \bbR \quad S^t(\tilde{\si})= -\log \frac{\d
\tilde{\bbP}^{\theta} (\theta \tilde{\si})}{\d \tilde{\bbP}
(\tilde{\si}) } \eeq We upgrade the function $w^t$ on $\Om_t$ to a
function on $\tilde{\Omega}_t$ as

\beq w^t(\tilde{\si})=w^t(\si) ,  \quad \tilde{\si}=(\si,i,j) \eeq
Our strategy will be to prove (Section \ref{sec: pos of S}) that
for some $u >0$

 \beq \label{eq: strict pos of S}
\int_{\tilde{\Om}_u} \d \tilde{\bbP}_u (\tilde{\si}) S^u(\tilde{\si}) >0
 \eeq
and then (Section \ref{sec: diff S and w}) that for all $t \geq 0$
 \beq
\int_{\tilde{\Om}_t} \d \tilde{\bbP}_t (\tilde{\si})
\big(S^t(\tilde{\si})-w^t(\tilde{\si})\big) \leq 0
  \eeq
which will lead to the conclusion that for a certain $u \in
\bbR^+$,
 \beq \int_\Om \d \bbP_{\rho} (\si)
w^u(\si) = \int_\Om \d \tilde{\bbP}_u (\tilde{\si})w^u(\tilde{\si})
>0
  \eeq
 where the first equality is checked by arguing as in \eqref{eq: check
 normalization}.
The converse statement is proven in Section \ref{sec: strict
implies A2}.
\subsubsection{Positivity of $S^t$} \label{sec: pos of S}
Looking back at the calculation \eqref{eq: path reversal}, one
immediately checks that for $t \geq 0$ and $\si \in \Om $, \beq
\label{eq: improved ft} \Tr \left [\frac{\rho_i}{\|\rho_i \|}
\caW_t(\si) \frac{(T\rho T)_j}{\| (T\rho T)_j\|} \right] =
\e^{w^t(\si)}\Tr \left [\frac{\rho_j}{\| \rho_j\|}
\caW^{\theta}_t(\theta \si)       \frac{(T\rho T)_i}{\| (T\rho
T)_i\|} \right] \eeq and hence \beq  \label{eq: improved def S}
S^t(\tilde{\si})= w^t(\tilde{\si}) - \log (\|\rho_j \|)+\log
(\|\rho_i \|), \qquad \tilde{\si}=(\si,i,j) \eeq

\noindent Note, using \eqref{def: S} and
$S^t(\tilde{\si})=-S^t(\theta_t \tilde{\si})$, that $S^t$
satisfies an exact fluctuation symmetry, for $t \geq 0$ and $\ka
\in \bbC$:

\beq \int_{\tilde{\Om}_t} \d \tilde{\bbP}_t (\tilde{\si}) \e^{-\ka S^t
(\tilde{\si})} = \int_{\tilde{\Om}_t} \d \tilde{\bbP}^{\theta}_t
(\tilde{\si}) \e^{-(1-\ka)S^t (\tilde{\si})} \eeq
Remark that $f: \mathbb{R} \rightarrow \mathbb{R}: x \rightarrow
e^{-x}+x-1 $ is positive for all $x$, increasing for $x \geq 0$
and decreasing for $x \leq 0$. A Chebyshev inequality with $\delta
>0$ yields
\begin{equation}\label{strict2}
\int_{\tilde{\Om}_t} \d \tilde{\bbP}_t(\tilde{\si}) S^t(\tilde{\si})=\int_{\tilde{\Om}_t} \d
\tilde{\bbP}_t(\tilde{\si}) (e^{-S^t}+S^t-1)(\tilde{\si}) \geq
(e^{-\delta}+\delta -1) \tilde{\bbP}_t(|S^t| \geq \delta)
\end{equation}
Rephrasing \eqref{cond: cycle nonzero ent}-\eqref{cond: closed
cycle}, there is for $u >0$, a $E \subset \Om_u$, and
one-dimensional projection $P \in \caB(\caH)$  such that

\beq \int_E \d \si \Tr \left[ P \caW_u(\si) P \right]>0  ,
\qquad    w^u( E)=:w \neq 0 \eeq

\noindent For any $k \in \bbN$, we construct \beq   \Om_{ku}
\supset E^k := \{\si_1 \si_2 \ldots \si_k ,\ | \,  \si_1,\ldots,
\si_k \in E \} \eeq where the notation $\si_1 \si_2$, and
consequently also $\si_1 \si_2   \ldots \si_k  $ was defined in
\eqref{glue in Om}. We have \beq w^t(E^k)=kw   , \qquad
\int_{E^k} \d \si \Tr \left[ P \caW_u(\si) P \right]>0   \eeq
Since $\rho$ is faithful, there are $i,j \in \{1,\ldots,\dim \caH
\}$ such that \beq \int_{E^k} \d \si \Tr \left[ \rho_i \caW_{ku}
(\si) \rho_j\right]
>0 \eeq

Since the function $S^u-w^u$ is bounded uniformly in $u \in
\bbR^+$ (this follows e.g. from \eqref{eq: improved def S}), one
can choose $k \in \bbN$ and $i,j \in \{1,\ldots,\dim \caH \}$ such
that  \beq   \Tr \left[ \rho_i \caW_{kt} (\si) \rho_j \right]= kw
+ \log  \| \rho_i \| -\log \| \rho_j \|
>0,   \qquad \si \in E^k \eeq
This proves that the last expression in \eqref{strict2} is not
zero (after replacing $t$ by $ku$).  Hence, \eqref{eq: strict pos
of S} is proven.

\subsubsection{Difference between $S^t$ and $w^t$} \label{sec: diff S and w}
Calculate for $t \geq 0$
 \begin{eqnarray}
 && \int_{\tilde{\Om}_t} \d \tilde{\bbP}_t(\tilde{\si})  \log \|\rho_j \|
= \sum_{i,j \in D} \int_{\Om_t} \d \si   \Tr \left[ \rho_i \caW_t(\si)  \frac{
(T\rho T)_j}{\| \rho_j \|} \right] \log \|
\rho_j \| \nonumber  \\
&=&   \Tr \left[ \rho  \e^{t \caL}  \log T \rho T \right] =\Tr \left[
\rho  \log  T\rho T \right]
 \end{eqnarray}
and
\begin{eqnarray}
&& \int_{\tilde{\Om}_t} \d \tilde{\bbP}(\tilde{\si})  \log \|\rho_i \| =
\sum_{i,j \in D} \int_{\Om_t} \d \si   \Tr \left[ \rho_i \caW_t(\si)  \frac{ T\rho_j T}{\| \rho_j \|}\right] \log \|\rho_i \| \nonumber\\
&=&\sum_{i \in D} \Tr \left[ \rho_i \e^{t \caL}1   \right] \log{ \|\rho_i \|}
= \Tr \left[ \rho \log{ \rho } \right]
 \end{eqnarray}
where we used $\rho \e^{t \caL}=\rho$ and $\e^{t \caL}1=1$.

 \noindent Hence one gets \beq \int_{\tilde{\Om}_t} \d
\tilde{\bbP}_t(\tilde{\si}) \big(\log((T\rho T)_i) - \log (\rho_j)
\big)=\Tr \left[ \rho (\log \rho -\log T\rho T) \right] \leq 0
\eeq where the last inequality follows from the nonnegativity of
the relative entropy.

\subsubsection{Strict positivity implies Assumption {\bf A2}} \label{sec: strict implies
A2}

We prove that {\bf A2} is a necessary condition for a nonzero
entropy production. First remark that \beq \label{eq: assume
nonzero ent} \int_{\Om_t} \d \bbP_{\rho}(\si) w^t(\si)
 \eeq
 is extensive in $t >0$. This follows from translation invariance
 (in $t$) of $w^t$ and stationarity of $\bbP_{\rho}$. Hence we can
 fix $t>0$ such that \beq \label{choose t for high
ent} \big| \int_{\Om_t} \d \bbP_{\rho}(\si) w^t(\si) \big|
> 2 \dim \caH \max_{\om,k} |\be_k \om | \eeq
Take $\si \in \Om_t$ satisfying $\caW_t(\si) \neq 0$. It follows
that one can split $t=\sum^3_{i=1} t_i$ and \beq \label{split of
si}\si =\tau_{3} \tau_2 \tau_{1}  , \qquad \tau_i \in \Om_{t_i} ,
\qquad  i=1,2,3 \eeq (again the notation \eqref{glue in Om} was
used) such that
\begin{enumerate}
\item{There is a one-dimensional projection
$P$ such that \beq \Tr \left[ P \caW_{t_2}(\tau_2) P
\right]>0 \eeq}
\item{\beq \label{eq: splitting
small} |\tau_1| \leq \dim \caH  ,\qquad |\tau_3| \leq \dim \caH \eeq    }
\end{enumerate}

\noindent Assume that {\bf A2} does not hold. It follows that  $
w^{t_2}(\tau_2)=0 $.

\noindent Hence, by \eqref{eq: splitting
small},
 \beq |w^t (\si) | = |w^{t_1} (\tau_1) +w^{t_3} (\tau_3)  |\leq
2 \dim  \caH \max_{\om,k} |\be_k \om | \eeq which is in obvious
contradiction with \eqref{choose t for high ent}.


\subsection{Proof of Proposition \ref{thm: green kubo}}

This proof is by now quite standard, it can be found e.g. in
\cite{lebowitzspohn1}.
We recall from \eqref{eq: derivatives converge} in Lemma \ref{lem:
analytic} that we can interchange the limit $t \uparrow \infty$
and differentiation of $\overline{\ka} \mapsto e(\overline{\ka})$.
By differentiating relation \eqref{eq: general ft} with respect to
$\ka_k$ and to $\be_{k'}$ in $\overline{\ka}=0$ and $ \beta_{k \in
K} =\be$, and interchanging limits and derivatives, we arrive at
the modified Green-Kubo relation:
\begin{equation} \label{modified green kubo}
L_{k,l}+ L^{\theta}_{k,l} = \beta \lim_{t \uparrow +\infty}
\frac{1}{t} \mathbb{E}_{\rho} [n^t_{k} n^t_{k'}]
\end{equation}
from which the other statements in Proposition \ref{thm: green kubo} easily follow.

\subsection{Proof of proposition \ref{conserv}}

Choose $v \in \caH$ and $\phi \in L^2(\Om_t)$ such that

\beq \label{choose eigenvectors} H_\sys v= m_S v   \qquad \forall
k \in K: \,  \mathbf{N}^t_k  \phi = m_k \phi     \qquad m_S, m_{k
\in K} \in \bbR \eeq

\noindent By the definition of $\mathbf{U}_t$, \beq
\label{calculating U} \mathbf{U}_t(v \otimes \phi) (\xi)
=\sum_{\si \subset \xi} \int_\Om \d \tau \, u_t( \si,\tau)v
\,\phi((\xi \setminus \si ) \cup \tau) , \quad \xi \in \Om_t \eeq

\noindent Using $[S,H_\sys]=0$, one checks that $u_t( \si,\tau)v $
either vanishes or
 \beq \label{change S energy} H_\sys  \big( u_t(
\si,\tau)v -v \big) = \sum_{\om ,k } \om
(|\si_{\om,k}|-|\tau_{\om,k}|) \big( u_t( \si,\tau)v -v \big) \eeq

\noindent By \eqref{choose eigenvectors}, it follows that
$\phi((\xi \setminus \si ) \cup \tau)=0$ in \eqref{calculating U},
unless for all $k \in K$ \beq \sum_{\om} \om \big(|\xi_{\om,k}|-
|\si_{\om,k}| + | \tau_{\om,k}| \big)=m_k \eeq Together with
\eqref{change S energy}, this implies \beq \mathbf{N}^t
\mathbf{U}_t(v \otimes \phi)= \mathbf{N}^t  (v \otimes \phi)=( m_S
+\sum_{k \in K} m_k ) (v \otimes \phi)\eeq Since the operators
$H_\sys, \mathbf{N}^t_{k \in K}$ mutually commute, vectors like $v
\otimes \phi$ as in \eqref{choose eigenvectors} furnish a complete
set of eigenvectors. This proves the proposition.

\subsection{Proof of Proposition \ref{current-current}}

By expanding the left-hand side of \eqref{currenteq} in a Dyson
expansion, as in \eqref{def: dyson}, one can evaluate the
derivatives, leading to
\begin{equation}
\frac{\partial^2}{\partial v_1 \partial v_2} \mathbb{E}_{\rho} [
n^{v_1}_{k}n^{v_2}_{k} ]\big|_{v_1=0,v_2=u}=
\sum_{\omega_1,\omega_2} \omega_1 \omega_2 \Tr \left[ \rho
\caJ_{\omega_1,k_1} e^{u \mathcal{L}}\caJ_{\omega_2,k_2}
(1)\right]
\end{equation}
 Putting $\rho = \rho_{\beta}$, yields
$\rho_{\beta}V_{\omega,k}=V_{\omega,k}\rho_{\beta}e^{-\beta_k
\omega}$. Now \eqref{currenteq} follows after some reshuffling,
using

\begin{equation}
V_{\omega,k}^*=V_{-\omega,k} ,\quad   \eta_k(\om)=\e^{-\be_k \om}
\eta_k(-\om) ,\quad \sum_{\omega} \omega \eta_k(\om)
V_{\omega,k}^*V_{\omega,k}=\caL_k(H_\sys)
\end{equation}

\renewcommand{\theequation}{A-\arabic{equation}}
  \setcounter{equation}{0}  
  \section*{APPENDIX}  

Let $\caA$ be the matrix algebra $M_n(\bbC)$ for some $n \in
\bbN$, and denote by $\caA^+$ its positive cone, i.e,  \beq \caA^+
=\{x^*x \,|\, x \in \caA \}\eeq An element $x \in \caA^+$ is
called strictly positive (notation: $x>0$) if it is invertible.
\begin{theorem}\label{app: perron}
Let $\La: \caA \to \caA$ be a completely positive
linear map, satisfying
\begin{equation} \label{cond: positivity}
   \Tr \left[ x \La y \right]> 0, \qquad x,y \in \caA^+, x \neq
   0, y \neq 0
\end{equation}
Then, $\La$ has a positive eigenvalue $\la$, such that if $\mu$ is
another eigenvalue, then $|\mu | <\la$. The eigenvector $v \in
\caA$ corresponding to $\la$ can be chosen strictly positive. The
eigenvalue $\la$ is simple, i.e., as a root of the characteristic
equation of $\La$ it has multiplicity $1$.
\end{theorem}

The theorem was proven almost in the above form in
\cite{evanshoegh}, (see Theorem 4.2 therein). We state (a
simplified version of) that theorem and we show that the above
statement follows from it. We call a positive map $\phi$ on $\caA$
irreducible if \beq  \forall x \neq 0,y \neq 0  \in \caA^+ ,
\exists k \in \bbN : \,   \Tr \left[ x \phi^k y \right]>0 \eeq
\begin{theorem}\label{app: evans1}
Let $\phi$ be a positive map  such that
\begin{enumerate}
\item{$\phi$ preserves the unit $1 \in \caA$:  $\phi(1)=1$}
\item{$\phi$ satisfies the two-positivity inequality:
\beq \label{two pos}\phi(x^*x) \geq \phi(x)^* \phi(x)    \textrm{
for all } x\in \caA  \eeq}
\item{ For all $k=1,2,\ldots $, $ \phi^k$ is irreducible}
\end{enumerate}
Then, $\phi$ has a positive, simple eigenvalue $\la$, such that if
$\mu$ is another eigenvalue, then $|\mu | <\la$. The eigenvector
$v \in \caA$ corresponding to $\la$ can be chosen strictly
positive.
\end{theorem}
Another theorem in \cite{evanshoegh} is (Theorem 2.4, combined
with the sentences following it):
\begin{theorem}\label{app: evans2}
Let $\phi$ be an irreducible positive linear map on $\caA$ and let $r$ be
the spectral radius \beq r:=\sup \{|c| \,\big| \, c \in \sp \phi
\} \eeq then there is a unique eigenvector $v \in  \caA^+$ with
eigenvalue $r$.
\end{theorem}

To prove Theorem \ref{app: perron}, we remark that $\La$ has the
same spectral properties as a well-chosen map $\phi$ that
satisfies the conditions of Theorem \ref{app: evans1}:\\ Since
$\La$ is irreducible, one can apply Theorem \ref{app: evans2} to
find an eigenvector $v$. Because of \ref{cond: positivity}, we
conclude that $v >0$. Let now the map $\phi$ be defined as \beq
\phi (x) =\frac{1}{r} v^{-1/2} \La(v^{1/2} x v^{1/2}) v^{-1/2} ,
\qquad x \in \caA \eeq It is clear that
\begin{enumerate}
\item{$\phi$ is completely positive and $\phi$ still satisfies \ref{cond: positivity}. }
\item{$\phi(1)=1$}
\item{$\sp \phi =\frac{1}{r}\sp \La$ and also the multiplicities of the eigenvalues are equal}
\end{enumerate}

Hence $\phi$ satisfies the conditions of Theorem \ref{app:
evans1}, since unity-preserving completely positive maps satisfy
the two-positivity inequality \eqref{two pos}. Theorem \ref{app:
perron} follows.

\section*{ Acknowledgments}

We thank Luc Bouten, Hans Maassen, Andr\'e Verbeure, Frank Redig
and Karel Neto\v cn\'y for stimulating discussions. We are
grateful to an unknown referee for suggesting numerous
improvements.

\bibliographystyle{myalpha}
\bibliography{langevinmay}
\end{document}